\documentstyle[11pt]{article}

\begin{document}


\newcommand{\al}{\alpha}
\newcommand{\bet}{\beta}
\newcommand{\ga}{\gamma}
\newcommand{\del}{\delta}
\newcommand{\ep}{\epsilon}
\newcommand{\epx}{\varepsilon}
\newcommand{\ze}{\zeta}
\newcommand{\th}{\theta}
\newcommand{\thx}{\vartheta}
\newcommand{\io}{\iota}
\newcommand{\la}{\lambda}
\newcommand{\ka}{\kappa}
\newcommand{\pix}{\varpi}
\newcommand{\rhx}{\varrho}
\newcommand{\si}{\sigma}
\newcommand{\six}{\varsigma}
\newcommand{\yp}{\upsilon}
\newcommand{\om}{\omega}
\newcommand{\phx}{\varphi}
\newcommand{\Ga}{\Gamma}
\newcommand{\De}{\Delta}
\newcommand{\Th}{\Theta}
\newcommand{\La}{\Lambda}
\newcommand{\Si}{\Sigma}
\newcommand{\Yp}{\Upsilon}
\newcommand{\Om}{\Omega}


\newcommand{\be}{\begin{eqnarray}}
\newcommand{\ee}{\end{eqnarray}}
\newcommand{\jt}{\tilde{J}}
\newcommand{\Ra}{\Rightarrow}
\newcommand{\lra}{\longrightarrow}
\newcommand{\pr}{\partial}
\newcommand{\ti}{\tilde}
\newcommand{\ng}{|0\rangle_{gh}}
\newcommand{\pj}{\prod J}
\newcommand{\pjt}{\prod\tilde{J}}
\newcommand{\prb}{\prod b}
\newcommand{\prc}{\prod c}
\newcommand{\bft}{|\tilde{\phi}>}
\newcommand{\bfj}{|\phi>}
\newcommand{\lan}{\langle}
\newcommand{\ran}{\rangle}
\newcommand{\bz}{\bar{z}}
\newcommand{\bJ}{\bar{J}}
\newcommand{\vacr}{|0\rangle}
\newcommand{\vacl}{\langle 0|}
\newcommand{\IFF}{\Longleftrightarrow}
\newcommand{\phr}{|phys\ran}
\newcommand{\phl}{\lan phys|}
\newcommand{\nonu}{\nonumber\\}
\newcommand{\tg}{\tilde{g}}
\newcommand{\tM}{\ti{M}}
\newcommand{\hd}{\hat{d}}
\newcommand{\hL}{\hat{L}}
\newcommand{\sir}{\si^\rho}
\newcommand{\pil}{\pi_{L'}}
\newcommand{\cg}{c_{\bar g}}

\newcommand{\ind}{\indent}
\newcommand{\np}{\newpage}
\newcommand{\hs}{\hspace*}
\newcommand{\vs}{\vspace*}
\newcommand{\nl}{\newline}
\newcommand{\bqu}{\begin{quotation}}
\newcommand{\equ}{\end{quotation}}
\newcommand{\bit}{\begin{itemize}}
\newcommand{\eit}{\end{itemize}}
\newcommand{\ben}{\begin{enumerate}}
\newcommand{\een}{\end{enumerate}}
\newcommand{\ul}{\underline}
\newcommand{\nn}{\nonumber}
\newcommand{\lef}{\left}
\newcommand{\rig}{\right}
\newcommand{\fra}{\twelvefrakh}
\newcommand{\Bb}{\twelvemsb}
\newcommand{\bT}{\bar{T}(\bz)}
\newcommand{\dagg}{^{\dagger}}
\newcommand{\qd}{\dot{q}}
\newcommand{\cP}{{\cal P}}
\newcommand{\hg}{\hat{g}}
\newcommand{\hh}{\hat{h}}
\newcommand{\hpg}{\hat{g}^\prime}
\newcommand{\htg}{\tilde{\hat{g}}^\prime}
\newcommand{\pri}{^\prime}
\newcommand{\lap}{\la^\prime}
\newcommand{\rhop}{\rho^\prime}
\newcommand{\Dgp}{\Delta_{g^\prime}^+}
\newcommand{\Dg}{\Delta_g^+}
\newcommand{\Pro}{\prod_{n=1}^\infty (1-q^n)}
\newcommand{\Pg}{P^+_{\hg}}
\newcommand{\Pgp}{P^+_{\hg\pri}}
\newcommand{\hmu}{\hat{\mu}}
\newcommand{\hnu}{\hat{\nu}}
\newcommand{\hrho}{\hat{\rho}}
\newcommand{\gp}{g^\prime}
\newcommand{\pp}{\prime\prime}
\newcommand{\CM}{\hat{C}(g',M')}
\newcommand{\CI}{\hat{C}(g',M^{\prime (1)})}
\newcommand{\CL}{\hat{C}(g',L')}
\newcommand{\HL}{\hat{H}^p (g',L')}
\newcommand{\HMI}{\hat{H}^{p+1}(g',M^{\prime (1)})}
\newcommand{\da}{\dagger}
\newcommand{\wro}{w^\rho}
\renewcommand{\Box}{\rule{2mm}{2mm}}

\newcommand{\NPB}[1]{Nucl. Phys. B  {#1} }
\newcommand{\IJMPA}[1]{Int. J. of Mod. Phys. A  {#1} }
\newcommand{\PLB}[1]{Phys. Lett. B {#1} }

\thispagestyle{empty}
\begin{flushright}
August, 1997
\\
HKS-NT-FR--97/1--SE\\
hep-th/9708100\\
\end{flushright}
\vs{5mm}

\begin{center}

{\huge{The semi-infinite cohomology of affine Lie algebras}} \\
\vspace{10 mm}
{\Large{Stephen Hwang}\footnote{Stephen.Hwang@hks.se}} \vspace{2mm}\\

\vspace{4 mm}

Karlstad University \\
S-651 88 Karlstad\\
Sweden

\vs{15mm}

{\bf Abstract }  \end{center}
We study the semi-infinite or BRST cohomology of affine Lie algebras in detail.
This cohomology is relevant in the BRST approach to gauged WZNW models.
Our main result is to prove necessary and sufficient conditions on ghost
numbers and
weights for non-trivial elements in the cohomology. In particular we prove the
existence of an infinite sequence of elements in the cohomology for
non-zero ghost
numbers. This will imply that the BRST approach to topological WZNW model
admits many
more states than a conventional coset construction. This conclusion also
applies to some
non-topological models.

 Our work will
also contain results on the structure of Verma modules over affine
 Lie algebras.
In particular, we generalize the results of Verma and
Bernstein-Gel'fand-Gel'fand,
for finite dimensional Lie algebras, on the structure and multiplicities of
Verma modules.

The present work
gives the theoretical basis of the explicit construction of the
elements in cohomology presented previously. Our analysis proves and makes
use of the close relationship between highest weight null-vectors and
elements of the cohomology.

\np
\setcounter{page}{1}
\section{Introduction and summary of results}

The present work studies the semi-infinite or BRST cohomology of affine Lie
algebras.
The motivation comes from the quantization of Wess-Zumino-Novikov-Witten
(WZNW) models. These models play an essential part in the understanding
 and classification of conformal field theories. The BRST symmetry
arises as a consequence of the gauging of a WZNW model w.r.t.
a subgroup \cite{KS}. The constraints associated
with this BRST symmetry are the generators of an affine Lie
algebra $g'=g_k\oplus \ti{g}_{\ti{k}}$. Here $g_k$ and $\ti{g}_{\ti{k}}$
correspond
to the same finite dimensional Lie algebra, but have different
central elements $k$ and $\ti{k}=-k-2c_{\bar{g}}$ (see section 2 for notation).
The latter affine Lie algebra corresponds to an {\it auxiliary}, and in
general non-unitary, WZNW model that arose in the derivation in \cite{KS}.
 The physical states in the gauged WZNW model are now given by the
non-trivial elements of the resulting BRST cohomology. In \cite{HR}
it was proved that the BRST approach was equivalent to the conventional
coset construction, so that the states were ghost-free and satisfied
the usual highest weight conditions w.r.t. the subalgebra $g_k$. The
conditions for this proof was that one selected a specific range of
representations for the auxiliary WZNW model. For the original
ungauged WZNW model the range of representations were assumed
to be the integrable ones.

In this work we will consider completely general highest weight representations
(an analagous treatment may be given for lowest weight
representions). The
motivation for this is that it may be that a more
general situation than in ref.\cite{HR} is the physically relevant
one.
Our analysis of the cohomology is most straightforwardly applied to the case
when the gauged subgroup coincides with the original group i.e. when we
have a topological WZNW model. But, as we will show, it also generalizes to
the most
important class of non-topological models, namely those in which the ungauged
WZNW model is unitary.

In \cite{HR2} the explicit construction of elements in the BRST
cohomology was considered. The procedure presented there for
obtaining these elements showed that they were
intimitely related to certain null-vectors. The key to the construction
 was to make a selection of null-vectors that generated
the states in the cohomology. It turned out that these null-vectors  are
the highest
weight vectors. Then by using the explicit form of highest weight
null-vectors given by
 Malikov, Feigin and Fuchs \cite{MFF}, the elements may be constructed.
Our work here may be seen as the theoretical basis of this
 construction. We will here prove that the procedure in \cite{HR2}
will always generate non-trivial states in the cohomology. We will also
prove that the ghost numbers that appeared in the construction are the {\it
only possible
ones}. The ghost numbers will be uniquely determined
by the representations of the algebras involved, and for fixed representations
only one value (and its negative) will occur. It is still an open question
whether the
construction provides all the possible states. We also lack a general
result on the
dimensionality of the cohomology.

The plan of the paper and its main results are the following. In section 2 we
 give basic definition and facts for affine algebras and associated
modules. In section 3 we discuss the structure of Verma
modules. This is important since our analysis of the cohomology
relies very heavily on this structure, in particular, on the embeddings of
Verma modules
into Verma modules.
We make extensive use of a technique due to Jantzen \cite{Ja} to
perturb the highest weight of a reducible Verma module to obtain
an irreducible one. This perturbation gives also a
filtration of modules in a given Verma module. Section 3 contains results
on the structure of Verma modules, which we have been unable to find in
the literature. The main results are Theorem 3.10 and
Theorem 3.11. These are generalizations  of results of Verma
\cite{V} and Bernstein, Gel'fand and Gel'fand
\cite{BGG},  respectively, for finite dimensional Lie algebras and of
Rocha-Caridi and
Wallach for affine Lie algebras with highest weights on Weyl orbits through
dominant weights.
The proof of Theorem 3.11 is almost identical to the proof of the finite
dimensional case given in
\cite{Dix}, Theorem 7.7.7 (which is used also in
\cite{RW1}). The proof of Theorem 3.10 only partly coincides with
\cite{RW1}, as the latter does not extend to the case of antidominant weights.

In section 4 we proceed to introduce the BRST formalism. Most
of the material
(except Lemma 4.2) is well-known. In particular, we recapitulate
a theorem due to Kugo and Ojima \cite{KO}.
This theorem will partly be used in the main
section, section 5. It is also conceptually important in understanding
the basic mechanism behind the
appearance of elements in the
BRST cohomology for non-zero ghost numbers, which we now explain.
The theorem, which applies only to
irreducible modules, states that elements in the cohomology form either
singlet or doublet ({\it singlet pair}) representations w.r.t the BRST
algebra. Furthermore,
elements that are trivial or outside the cohomology form so-called {\it
quartets} in
the terminology of \cite{KO} i.e. sets four states, in which two of the
elements are BRST
exact. In order  to obtain an irreducible module,
 we use a trick due to Jantzen, to perturb a
reducible module into an irreducible one. In the irreducible case one may
prove
(Corollary 5.2),  that only ghost-free highest weight
 states are BRST non-trivial. As the perturbation is taken to zero and the
module
becomes reducible, certain quartets will
evolve into singlet pairs in the
following way. Two of states of the quartet will remain in the irreducible
module and will then form a singlet pair in this module. The two other states
will become null-states. One of the main results in this paper (Theorem 5.12)
is the determination of the relevant null-states.
This theorem gives the necessary and sufficient
conditions on the null-states to be part of a quartet, that will contain a
singlet
pair as the perturbation is set to zero. The implications of the theorem is
exploited in
Theorems 5.14 and 5.15, which give the necessary and sufficient conditions
on the
ghost-numbers and weights for which the cohomology is non-trivial. In
particular in
Theorem 5.15 a sequence of non-trivial BRST invariant states is proved to
exist. This
sequence is exactly the one for which the construction has been given in
\cite{HR2}. The ghost numbers appearing are $\pm p$, where
$p=l(\ti{\la})-l(\la)$ and $l(\la)$ is the length of a Weyl
transformation associated with $\la$ (see section 3). This means that for given
highest weights $\la$ and  $\ti{\la}$ of the original and auxiliary sectors,
$|p|$ is fixed to exactly one value. By Theorem 5.14 these ghost numbers
and weights are
the only non-trivial ones.

Let us also address the question of how the embedding of $g$ into a larger
algebra may affect our results. As our approach relies on the use of
null-vectors, the
crucial question is what happens to the relevant null-vectors as g is embedded.
If the null-vector w.r.t. $g$ will cease to be null in the larger algebra, then
the entire quartet, to which the vector belongs for non-zero perturbation,
 will remain a quartet as
Jantzen's perturbation is set to zero. Thus the corresponding elements in the
cohomology of $g$ will now be exact. In addition, many more elements may
disappear from the cohomology group. This is most evident from the
construction in \cite{HR2}, where one used non-trivial states at ghost
number $p-1$ ($p>0$) to construct a BRST non-trivial
element of ghost number $p$. In the extreme case the module over the larger
algebra is irreducible and all elements, except the one at zero ghost number,
will disappear.

There is one case in which the embedding will be straightforward. This will
happen
when we select integrable representations of the larger algebra. In this case
it is known \cite{KP} that the irreducible module over the larger algebra
is completely
reducible w.r.t. to any subalgebra. Hence, the results given here
generalize directly.
This was the situation analyzed in \cite{HR}. Corollary 5.11 proves
that the solutions given in \cite{HR} for a selected range of representations
of the auxiliary sector, are in fact the {\it unique} solutions for zero ghost
number for any selection of representations of the auxiliary sector.

The existence of extra elements in the cohomology, which have non-zero
ghost numbers,
implies that the BRST approach to WZNW models is different from the conventional
coset approach. This applies to the topological case, but also to the
non-topological case, at least when we take integrable representations of
the original
algebra. The r\^{o}le of these extra states is at this point unclear. It may be
that their appearance will lead to inconsistencies. One may avoid the
states by selecting an
appropriate range of representations for the auxiliary sector. Then only ghost free states
will appear in the cohomology. This was the situation treated in ref
\cite{HR}.  It may on the
other hand be that the extra states are a new and important part of the
quantization of WZNW
models. In the latter case one may expect that the extra states will be
needed to ensure
S-matrix unitarity and hence will appear as poles in scattering amplitudes.

\section{Preliminaries}
Let $\bar{g}$ be a simple finite dimensional Lie algebra of rank $r$. We denote
by $g_k$ the corresponding affine Lie algebra of level $k$. The set of roots of
$\bar{g}$ and $g$ are $\bar{\al}\in\bar{\Delta}$ and $\al\in\Delta$,
respectively. The highest root of $\bar{g}$ is denoted $\bar{\psi}$ and its
length is taken to be one. The restriction to positive roots are denoted by
$\bar{\Delta}^+$, $\Delta^+$ and to simple roots by $\bar{\Delta}^s$,
$\Delta^s$. The weight and root lattices of $\bar{g}$ and $g$ are denoted by
$\bar{\Gamma}_w,\bar{\Gamma}_r,\Gamma_w$ and $\Gamma_r$. $\Gamma_r^+$ is the
lattice generated by positive roots. Let $\Gamma_w^+$ be the set of dominant
weights, $\Gamma_w^+=\{\la\in\Gamma_w\ |\ \al_i\cdot\la\geq 0 \mbox{ for
}\al_i\in\Delta^s\}$. Let $\Gamma_w^f=\{\la_i\in\Gamma_w^+\ |\
{2\la_i\cdot\al_j\over (\al_j)^2}=\delta_{ij} \mbox{ for } \al_j\in\Delta ^s\}$
be the set of fundamental weights. Here $\la_i\cdot\al_j$ denotes the
invariant scalar product on $g$ and $(\al_j)^2=\al_j\cdot\al_j$. Define $\rho$
as twice the sum of fundamental weights of $g$. $\bar{\rho}$ is the
corresponding
sum for $\bar{g}$. $\rho$ satisfies $\rho\cdot\al_i=(\al_i)^2$,
$\al_i\in\De^s$. We define
the set of antidominant weights
$\Gamma_w^-=\{\la\in\Gamma_w|\al_i\cdot(\la+\rho/2)\leq 0 \mbox{ for
}\al_i\in\Delta^s\}$. A weight $\mu\in\Ga_w$ is said to be singular if it
is orthogonal to
at least one positive root and is said to be regular otherwise.

The Weyl group $W$ of $g$ is the set of transformations on $\Gamma_w$ generated
by the simple reflections
\be
\sigma_i(\la)=\la-{2\la\cdot\al_i\over(\al_i)^2}\al_i\hs{5mm}\al_i\in\Delta^s.
\ee
The length $l(w)$ of $w\in W$ is the minimal number of simple reflections that
give $w$. We also define the $\rho-$centered reflections
$\sigma_i^\rho(\la)=\si(\la+\rho/2)-\rho/2$. Similarly we write $w^\rho(\la)$
for a general $\rho$-centered Weyl transformation. We define an ordering
between weights. Let $\mu,\nu\in \Gamma_w$ be such that
$\mu-\nu\in\Gamma_r^+$. We then write $\mu\geq\nu$. If
$\mu-\nu\in\Gamma_r^+/\{0\}$, then this is denoted by $\mu>\nu$. Two
weights $\la$ and
$\mu$ are said to be on the same Weyl orbit if there exists $w\in W$ such that
$\mu=w(\la)$. Similarly, they are said to be on the same $\rho$-centered
Weyl orbit if
$\mu=w^\rho(\la)$.

We make a triangular decomposition of $g$, $g=n_-\oplus h\oplus n_+$. We
will use the notation $e_\al$ for the
generators of $n_+$, $f_\al$ for those of $n_-$ and $h_i$, $i=1,\ldots,r+2$
for the generators of the Cartan subalgebra $h$. $h_i,\ i=2,\ldots, r+1$ span
$\bar{h}$, $h_{1}$ is a central element of $g$ with eigenvalue $k/2$ and
$h_{0}$ is a derivation. We have a corresponding decomposition of
${\cal U}(g)$, the universal enveloping algebra of $g$, as ${\cal U}(g)={\cal
U}(n_-)\otimes{\cal U}(h)\otimes{\cal U}(n_+)$.

 Let $M(\la)$ denote the highest weight Verma module over $g$ of highest
weight $\la$. The module is generated by a highest weight primary vector
$v_{0\la}$ satisfying
$$
e_\al v_{0\la}=0\nonumber$$
\begin{equation}h_i v_{0\la}=\la_i v_{0\la} \hs{5mm} h_i\in h.
\label{prim}
\end{equation}
$M(\la)$ admits a weight decomposition
$$
M(\la)=\bigoplus_{\eta\in\Gamma_r^+}M_\eta(\la).
$$
Vectors in $M_\nu(\la)$ will be called weight vectors of degree $\nu$
and their weights differ from the highest weight by $\nu$. We consider
throughout only vectors $v\in M_\eta(\la)$ with
dim$M_\eta(\la)<\infty$. The dimension of $M_\nu(\la)$ is $P(\nu)$, which is
the number of ways $\nu$ may be written as a linear combination of positive
roots with non-negative coefficients.  Let $M'(\la)$ be the proper
maximal submodule of $M(\la)$. Then $M(\la)/M'(\la)$ is irreducible and
isomorphic to the unique irreducible $g-$module $L(\la)$.

Define a Hermitean form $\lan ..|..\ran$ as the mapping from $M(\la)\times
M(\la)$ to the
complex numbers by
$$\lan v_{0\la}|v_{0\la}\ran=1$$
\begin{equation}\lan w_\la|uv_\la\ran =\lan u^\dagger
w_\la|v_\la\ran ,\label{hermf}\end{equation}
 where $u\in{\cal U}(g)$ and $(\
)^\dagger$ denotes the Hermite conjugation defined by
$e_\al^\dagger=f_\al,\ f_\al^\dagger=e_\al, \ h_i^\dagger=h_i$.
For $v_\eta,w_\mu\in
M_\eta(\la)$ we clearly have
$\lan w_\mu|v_\eta\ran=0$ for $\eta\neq \mu$. If $\eta=\mu$, then
$F(\la)_\eta=\lan w_\eta|v_\eta\ran$ may be viewed as a $P(\eta)\times P(\eta)$
matrix, whose entries are polynomials in $\la$. The determinant of
$F(\la)_\eta$ is given by the Kac-Kazhdan formula \cite{KK}
\be
\mbox{det }
F(\la)_\eta=\mbox{const.}\prod_{\al\in\Delta^+}\prod_{n=1}^\infty\left[(\la
+\rho/2)\cdot\al-\frac{n}{2}\al^2\right]^{P(\eta-n\al)}\label{kkdet}\ee
where roots $\al\in\Delta^+$ are taken with their multiplicities and
$P(\eta)=0$ if
$\eta\not\in\Gamma^+$. The zeros of the determinant are associated with highest
weight vectors $v_\mu$ that occur in $M(\la)$ (see the following section). From
eq.(\ref{kkdet}) one may infer that $\mu=\la-n\al$, which implies that the
Verma module $M(\mu)$ is a submodule of $M(\la)$. $M(\la)$ is irreducible if
and only if there does not exist $n\in{\cal Z}$ and $\al\in\Delta^+$ such that
\be
(\la+\rho/2)\cdot\al-\frac{n}{2}\al^2=0.\label{kkeq}
\ee
Notice that this equation will for any imaginary root $\al$ (i.e.
$\al^2=0$) be equivalent
to the condition $k=-c_{\bar{g}}$, where $c_{\bar{g}}$ is the quadratic
Casimir of the
adjoint representation of $\bar{g}$.

\section{Structure of embeddings of Verma modules}
\setcounter{equation}{0}
If the Kac-Kazhdan equation (\ref{kkeq}) has non-trivial solutions for a given
module $M(\la)$, then there will exist Verma modules $M(\mu)$ that are
submodules
of $M(\la)$. This implies the existence of a
$g$-homomorphism, $\phi\in\mbox{Hom}_g\left(M(\mu),M(\la)\right)$,
such that $M(\mu)\stackrel{\phi}{\rightarrow} M(\la)$. We will in this
section and throughout
the rest of this paper assume $k\neq -c_{\bar{g}}$, so that solutions to
eq.(\ref{kkeq}) only occur for real roots $\al$. The structure of embeddings
is most clearly depicted through a filtration due to Jantzen
\cite{Ja}. Introduce $z=\sum_{\la\in\Gamma_w^f}z_\la\la$, where
$z_\la$ are non-zero complex numbers. Consider the one-parameter family of
weights $\la_\ep=\la+\ep z$. If $\la$ is a weight of a reducible module
$M(\la)$ or $M^{\ast}(\la)$ and $z_i\neq 0$, then for $0<|\ep|\ll1$,
$M(\la_\ep)$ and $ M^\ast(\la_\ep)$ are irreducible. We now define a
filtration  \be M(\la_\ep)\supset M^{(1)}(\la_\ep)\supset
M^{(2)}(\la_\ep)\supset\ldots \ee by  \be
M^{(n)}(\la_\ep)=\left\{v\in M(\la_\ep)\ \mid\ \lan w^\ast|v\ran
\mbox{ is divisable by
}\ep^n\mbox{ for any }w^\ast\in M^\ast(\la_\ep)\right\}.
\ee
We will often write $M_\ep$ for $M(\la_\ep)$ etc. for $M^{(n)}$.
If $v=uv_{0\la}, u\in{\cal U}(g)$, then we write $v_\ep
=uv_{0\la_\ep}$.
In the dual case one may define a corresponding filtration. In the limit
$\ep\rightarrow 0$ this induces a filtration of modules in $M(\la)$
\be
M(\la)\supset M^{(1)}(\la)\supset M^{(2)}(\la)\supset\ldots
\ee
Note that Jantzen's filtration is hereditary: Let $M(\mu)\in M^{(s)}(\la)$ and
$M(\nu)\in M^{(t)}(\mu)$. Then $M(\nu)\in M^{(s+t)}(\la)$.

Any irreducible subquotient of a $g$-module $M(\la)$ is isomorphic to an
irreducible $g$-module $L(\mu)$, $\la-\mu\in \Ga_r^+$.
Denote by $(M(\la):L(\mu))$ the multiplicity of $L(\mu)$ in $M(\la)$.
$M^{(1)}(\la)$ is the maximal
proper submodule of $M(\la)$ and hence $M(\la)/M^{(1)}(\la)$ is isomorphic
to the
irreducible module $L(\la)$. We will call the vectors in $M^{(1)}$ null-vectors
of $M(\la)$. We define $\pi_L$ to be the projection
$M(\la)\stackrel{\pi_L}{\longrightarrow}L(\la)$.

The submodules
of a given Verma module are generally not all of Verma type. It is convinient
to introduce the notion of {\it primitive} vectors. Let $V$ be a $g$-module.
A vector
$v_\la\in V$ is said to be
primitive if there exists a submodule $U$ of $V$ such that
\be
v\not\in U,\hs{5mm} xv\in U\hs{2mm}\mbox{ for any }x\in n_+.
\ee
$\la$ is called a primitive weight. Highest weight vectors are clearly
primitive, but in general they do not exhaust all primitive vectors, even
in the case of finite dimensional algebras, as was first noted in \cite{BGG}.
In fact, there may be infinitely many more primitive vectors than highest
weight vectors (see \cite{CD} for an example for finite dimensional algebras).
Any module $V$ is
generated by its primitive weights as a $g$-module. We will call a module
which is
generated by acting freely with ${\cal U}(n_-)$ on a primitive vector,
which is not of
highest weight type, a {\it Bernstein-Gel'fand} (BG) module. The
corresponding primitive
vector will be called a {\it Bernstein-Gel'fand} primitive vector.

Although every zero in the determinant eq.(\ref{kkdet}), i.e. every
$(\al,n)$ for
which the Kac-Kazhdan eq.(\ref{kkeq}) is satisfied, corresponds to a highest
weight vector in $M(\la)$ (cf. Proposition 3.8), the converse is in general not
true. For a given $\la$ there are usually more
highest weight vectors than
solutions $(\al,n)$. Let $\mbox{Hom}_g(M(\mu), M(\la))\neq 0$ for a pair
$(\al,n)$ in eq.(\ref{kkeq}) with $\al$ real i.e. $\mu=\la-n\al,\ n\geq1\mbox{
and } \al\in\Delta^+\cap \Delta^R$, where $\Delta^R$ is the set of real roots.
Then we may write
\be
\mu=\si^\rho_\al(\la)<\la.\label{sr}
\ee
The inequality ensures that a solution to eq.(\ref{kkeq}) exists. In the form
eq.(\ref{sr}) it is clear that by iteration, we will find new
highest weight vectors not given by solutions to the Kac-Kazhdan equation for
$\la$. It also follows that $M(\la)$ is irreducible if and only if $\la$ is
antidominant. Notice that this requires $k<-c_{\bar g}$.

Let us proceed to give a more precise classification of highest weight vectors
in $M(\la)$ in terms of Weyl transformations. Define the Bruhat ordering on
$W$. Let $w,w'\in W$. We write $w'\rightarrow w$ if there exists
$\al\in\Delta^+\cap \Delta^R$, such
that $w=\si_\al w'$ and $l(w)=l(w')+1$. We write $w'\prec w$ if there are
$w_0,w_1,\ldots,w_p\in W$ such that $w'=w_p\rightarrow w_{p-1}\rightarrow
\ldots\rightarrow w_1\rightarrow
w_0 =w$. It may be shown that $w' \prec w$ if and only if the reduced
expressions
$w'=\si_{j_1}\ldots\si_{j_p}$ and $w=\si_{i_1}\ldots\si_{i_q}$
are such that $(j_1,\ldots ,j_p)$ is obtained by deleting $q-p$ elements
from $(i_1,\ldots,i_q)$.

By combining Theorem 4.2 in \cite{KK} with eq.(\ref{sr}) we have the following.
\vs{5mm}\\
{\sc Theorem 3.1.} A Verma module $M(\la)$ contains an irreducible subquotient
$L(\mu)$ if and only if the following condition is satisfied:
\bqu
\hspace*{-7mm}(*)\ $\la=\mu$, or there exists a sequence of positive roots
$\al_1,\al_2,\ldots,\al_k$ and a sequence of weights
$\la=\mu_1,\mu_2,\ldots,\mu_k,\mu_{k+1}=\mu$ such that
$\mu_{i+1}=\sir_{\al_i}(\mu_i)<\mu_i$ for $i=1,2,\ldots, k$.
\equ
{\sc Lemma 3.2.} Let $\mu\in\Gamma_w$. Then there exists
$w\in W$ and a unique $\la+\rho/2\in\Ga_w^+\ (k> -\cg)$ or
$\la\in\Ga_w^-\ (k<-\cg )$ such that
$\mu=\wro(\la)=\sir_{i_n}\sir_{i_{n-1}}\ldots\sir_{i_1}(\la)$, where
$i_1,\ldots,i_n$ denote the simple roots $\al_{i_1},\ldots,\al_{i_n}$
 with  \bqu
\hs{-7mm}(**)\ $\mu=\la$, or $\mu\neq\la$ and
$\sir_{i_{p+1}}\sir_{i_p}\ldots\sir_{i_1}(\la)< \sir_{i_p}\ldots
\sir_{i_1}(\la)\
(k> -\cg)$ or \\ $\sir_{i_{p+1}}\sir_{i_p}\ldots\sir_{i_1}(\la)>
\sir_{i_p}\ldots\sir_{i_1}(\la)\ (k< -\cg)$, \ $p=1,2,\ldots, n-1$. \equ
\vs{5mm}
{\sc Proof.} Consider $k<-\cg $. For $\mu\in\Ga_w^-$ the lemma is trivial
($w=1$).
Let $\mu=\mu_1\not\in\Ga_w^-$. Then there exists $\al_1\in\De^s$ such that
$n_1=(2\mu_1+\rho)\cdot\al_1/\al_1^2\in{\cal N}=1,2,3,\ldots.$. This
implies that
$\mu_2=\sir_{\al_1}(\mu_1)$ satisfies $\mu_2<\mu_1$. Let
$\la+\rho/2\in\Ga_w^-$ be such
that $(\mu_2-\la)^2\geq 0$ (which is always possible, as can be seen by an
explicit parametrization of the weights). We have
$(\mu_2-\la)^2=(\mu_1-\la)^2+n_1(2\la+\rho)\cdot \al_1$ and, therefore,
$(\mu_2-\la)^2<(\mu_1-\la)^2$. If $\mu_2\not\in\Ga_w^-$ we can continue this
process. We get a sequence of weights $\mu_1=\mu,\mu_2,\dots,\mu_r$
with $(\mu_{p+1}-\la)^2<(\mu_p-\la)^2$ and
$\mu_{p+1}=\sir_{\al_p}(\mu_p)<\mu_p$,
$p=1,\ldots,r-1$. This sequence must terminate after a finite number of steps,
since $(\mu_{r}-\la)^2\geq 0$ from $(\mu_2-\la)^2\geq 0$. But this can only
happen if the last weight $\mu_r$ of the sequence satisfies
$\al_i\cdot(2\mu_r+\rho)\leq 0$ for all $\al_i\in\De^s$ i.e. $\mu_r\in\Ga_w^-$.
We now prove the uniqueness.  Assume $w,w'\in W$ and $\la,\la'\in \Ga_w^+$ such
that $\mu=\wro(\la)=w^{\prime\rho}(\la')$. Then
$\la=w^{-1\rho}w^{\prime\rho}(\la')$.  This implies $\la=\la'$, as follows by an
adaption of \cite{Dix}, Lemma A in section 13.2, to the present case. The case
$k> -\cg$ is proved in a completely analogous fashion.$\Box$\vs{5mm}\\
{\sc Lemma 3.3.} Let $\mu$ and $\la$ be as in Lemma 3.2 and
$\mu_0=\la,\ \mu_1=\sir_{i_1}(\mu_0),\
\mu_2=\sir_{i_2}(\mu_1),\ldots,\mu_n=\sir_{i_n}(\mu_{n-1})=\mu$, where
$\sir_{i_k}$,
$k=1,2,\ldots,n$ are simple reflections satisfying (**). Then for
$k> -\cg$, Hom$_g\left(M(\mu_p),M(\mu_{p-1})\right)\neq 0,\ p=1,2,\ldots,n$ and
for $k<-\cg$, Hom$_g\left(M(\mu_{p-1}),M(\mu_p)\right)\neq 0,\ p=1,2,\ldots,n$.
\vs{5mm}\\
{\sc Proof.} The proof is by explicit construction. Consider e.g. $k<-\cg$ and
$\mu_p=\sir_{i_p}(\mu_{p-1})$. We take the $sl_2$ subalgebra generated by
$e_{i_p},\ f_{i_p}$ and $h_{i_p}$ satisfying $[f_{i_p},e_{i_p}]=h_{i_p}$ and
$[h_{i_p},f_{i_p}]=f_{i_p}$. Let $v_{\mu_p}$ be the highest weight vector that
generates $M(\mu_p)$ and $h_{i_p}v_{\mu_p}=\mu_{i_p}v_{\mu_p}$. Then it is
straightforward to check that $v_{\mu_{p-1}}=(f_{i_p})^{2\mu_{i_p}+1}v_{\mu_p}$
is a highest weight vector and it will generate a submodule isomorphic to
$M(\mu_{p-1})$. Hence, Hom$_g\left(M(\mu_{p-1}),M(\mu_p)\right)\neq 0$.
$\Box$\vs{5mm}\\
By Theorem 3.1, Lemma 3.2 and Lemma 3.3 and we have the following:\vs{5mm}\\
{\sc Proposition 3.4.} Let $\mu\in\Ga_w$. Then there exists a unique
$\la+\rho/2\in\Ga_w^+\ (k>-\cg )$, or $\la\in\Ga_w^-\ (k<-\cg)$, such that
Hom$_g\left(M(\mu),M(\la)\right)\neq 0\ (k>-\cg)$, or
Hom$_g(M(\la),$$M(\mu))\neq 0,\ (k<-\cg)$. Furthermore, if
$\nu\in\Ga_w$ and Hom$_g\left(M(\mu),M(\nu)\right)\neq 0$, then \\
$$[\mbox{dimHom}_g\left(M(\mu),M(\la)\right)][\mbox{dimHom}_g
\left(M(\nu),M(\la)\right)]\neq
0\mbox{\  for }k>-\cg\mbox{ or }$$
$$[\mbox{dimHom}_g\left(M(\la),M(\mu)\right)][\mbox{dimHom}_g
\left(M(\la),M(\nu)\right)]\neq
0\mbox{\  for }k<-\cg.$$ \vs{5mm}\\
{\sc Lemma 3.5.} Let $\la+\rho/2\in\Ga_w^+\ (k>-\cg )$ or $\la\in\Ga_w^-\
(k<-\cg)$, $w\in W$ and $\al\in\De^+\cap \Delta^R$.
Then\vs{2mm}\\
\hs{5mm}(i)  $\sir_\al\wro(\la)<\wro(\la)$ $\Rightarrow$
$l(\si_\al w)>l(w)$ for
$k>-\cg $ or $l(\si_\al w)<l(w)$ for $k<-\cg $.
\hs{4mm} (ii)  $l(\si_\al w)>l(w)$ for
$k>-\cg $ or $l(\si_\al w)<l(w)$ for $k<-\cg $ $\Rightarrow$
$\sir_\al\wro(\la)\leq\wro(\la)$
 \vs{5mm}\\
{\sc Proof.}
The proof of (i) is identical to that of Lemma 7.7.2 (ii) in \cite{Dix} (cf
\cite{RW1},
Lemma 8.2). Note that in the proof of Lemma 7.7.2 in \cite{Dix},
$\la\in\Ga_w^+$ is assumed. The
weaker condition on $\la$, assumed in our case, does not affect (i).
We prove (ii) for $k>-\cg $. We
have $$\sir_\al\wro(\la)=\wro(\la)-n\al.$$
Here $n={(2\wro(\la)+\rho)\cdot\al\over\al^2}\in{\cal
Z}$ If $n<0$ then $\sir_\al\wro(\la)>\wro(\la)$. By (i), we get $l(\si_\al
w)<l(w)$
which is a contradiction. Hence, $n=0,1,2,\ldots$ and (ii) follows.
The proof for $k<-\cg$ is
analogous. $\Box$\vs{5mm}\\
The following two lemmas are direct generalizations of \cite{Dix},
Lemma 7.7.4 and Lemma 7.7.5
(cf. \cite{RW1}, Lemma 8.4 and Lemma 8.5).\vs{5mm}\\
{\sc Lemma 3.6.} Let $w_1,w_2\in W$, $\ga\in \De^+\cap\De^R$ and
$\al\in\De^s$, with $\ga\neq \al$.
The following conditions are equivalent:\vs{2mm}\\
\hs{5mm}(i) $\si_\al w_1\stackrel{\al}{\longleftarrow}w_1$ and $\si_\al w_1
\stackrel{\ga}{\longleftarrow}w_2$\\
\hs{4mm}(ii) $w_2\stackrel{\al}{\longleftarrow}\si_\al w_2$ and $w_1
\stackrel{\si_\al(\ga)}{\longleftarrow}\si_\al w_2.$\vs{5mm}\\
{\sc Lemma 3.7.} Let $w\in W$ and  $\ga\in \De^+\cap\De^R$ be such that
$l(w)>l(\si_\ga w)$.
Then $w\succ \si_\ga w$.\vs{5mm}\\
We proceed to obtain results on the $g$-homomorphisms $M(\nu)\rightarrow
M(\mu)$.
First we have the following:\vs{5mm}\\
{\sc Proposition 3.8}. (cf. \cite{Dix}. Lemma 7.6.11). Let $\nu\in\Ga_w,
\al\in\De^+,\mu=\sir_\al(\nu)$. Assume $\mu\leq\nu$. Then
Hom$_g\left(M(\mu),M(\nu)\right)\neq 0$.\vs{5mm}\\
{\sc Proof.} The proof is essentially the same as in \cite{Dix}.
The case $\mu=\nu$ is trivial, so
we assume $\mu<\nu$. We consider only $k>-\cg $ as the the case $k<-\cg $
is analogous.
By Lemma 3.2 there
exists $w\in W$ and $\la'+\rho/2\in\Ga_w^+$ such that $\nu=\wro(\la')$. Let
$w=\si_{\al_n}\ldots \si_{\al_1}$ be a reduced expression of $w$ in terms of
simple reflexions and  $$\nu_0=\la',\ \nu_1=\sir_{\al_1}(\nu_0),\
\nu_2=\sir_{\al_2}(\nu_1),\ldots ,\nu_n=\sir_{\al_n}(\nu_{n-1})=\nu$$
$$\mu_0=\la,\ \mu_1=\sir_{\al_1}(\mu_0),\ \mu_2=\sir_{\al_2}(\mu_1),\ldots
,\mu_n=\sir_{\al_n}(\mu_{n-1})=\mu.$$
Then $\nu_0=w^{\prime\rho}(\mu_0)$ for some $w'\in W$ (from
$\nu_0=w^{-1\rho}(\nu)=w^{-1\rho}\sir_\al(\mu))$ and $\mu_0+\rho/2\in\Ga_w^+$,
hence $\mu_0-\nu_0\in\Ga_r^+$. On the other hand, $\mu_n-\nu_n=-m\al,\ m>0$.
Since the same element of $W$ transforms $\mu$ and $\nu$ into $\mu_p$ and
$\mu_p$, respectively, $p=0,1,2,\ldots, n$, $\mu_p$ is transformed from
$\nu_p$ by a reflexion $\sir_{\ga_p}$ ($\ga_p\in\De^+$), hence
$\mu_p-\nu_p\in\Ga_r^+$ or $\nu_p-\mu_p\in\Ga_r^+$. Hence, there exists a
smallest integer $k$ such that $\mu_k-\nu_k\in\Ga_r^+$ and
$\mu_{k+1}-\nu_{k+1}\in -\Ga_r^+$.
 Now $\mu_k-\nu_k=\sir_{\al_{k+1}}(\mu_{k+1}-\nu_{k+1})$. Since
$\mu_{k+1}-\nu_{k+1}$
is proportional to $\ga_{k+1}$, it can be seen that
$\si_{\al_{k+1}}(\ga_{k+1})\in\De^-$. Hence, $\ga_{k+1}=\al_{k+1}$ (since
$\si_{\al_{k+1}}$ permutes all positive roots except $\al_{k+1}$). The
relations $\mu_{k+1}-\nu_{k+1}\in -\Ga_r^+$ and
$\mu_{k+1}=\sir_{\al_{k+1}}(\nu_{k+1})$ imply
Hom$_g\left(M(\mu_{k+1}),M(\nu_{k+1})\right)\neq 0$ (Lemma 3.3). On the
other hand $M(\mu_{k+2})=M(\sir_{k+2}(\mu_{k+1}))$ so that
Hom$_g\left(M(\mu_{k+2}),M(\mu_{k+1})\right)\neq 0$. Hence,
Hom$_g\left(M(\mu_{k+2}),M(\sir_{\al_{k+2}}(\nu_{k+1}))\right)=
$Hom$_g\left(M(\mu_{k+2}),M(\nu_{k+2})\right)\neq 0$. Continuing this step
by step we arrive at Hom$_g\left(M(\mu),M(\nu)\right)\neq 0.$$\ \Box$\vs{5mm}\\
As a corollary to this proposition we can generalize results obtained by
\cite{V} and
\cite{BGG} for finite dimensional Lie algebras.\vs{5mm}\\
{\sc Corollary 3.9}. A necessary and sufficient condition for $M(\mu)$ to be a
submodule of $M(\nu)$ is that the condition (*) in Theorem 3.1 is
satisfied.\vs{5mm}\\
Note here the following. Firstly, Theorem 3.1 and Corollary 3.9 imply that
a BG module
$V(\mu)$ is a submodule of $M(\la)$ if and only if $(M(\la):L(\mu))\geq 2$.
Secondly, if
a BG module $V(\mu)\subset M(\la)$, then there exists a $g$-homomorphism
$\phi_{VM}$ such
that
$V(\mu)\stackrel{\phi_{VM}}{\rightarrow}M(\mu)\subset M(\la)$.

We are now ready to formulate one of the main results of this
section namely the dimension of the $g$-homomorphisms $M(\mu)\rightarrow
M(\nu)$.
This result generalizes the result of Verma \cite{V} for finite
dimensional Lie algebras and Rocha-Caridi, Wallach \cite{RW1} for
representations
with highest weights on Weyl orbits through dominant weights.\vs{5mm}\\
{\sc Theorem 3.10.} Let $\mu,\nu\in\Ga_w$. Then
dimHom$_g\left(M(\mu),M(\nu)\right)\leq 1$.\vs{5mm}\\
{\sc Proof.} We consider the cases $k>-\cg $
and $k<-\cg $ separately.\vs{2mm}\\
$k>-\cg $: By Proposition 3.4 it is sufficient to prove that dimHom$_g
\left(M(\mu),M(\la)\right)\leq 1$, where $\mu=\wro(\la),\ \la+\rho/2\in\Ga_w^+$.
The proof is then similar to that of \cite{RW1}, Lemma 8.14, using induction on $l(w)$.
We only
sketch it. For $l(w)=0$ the theorem is trivial. Assume it to be true for
$l(w)<p$. Consider
$l(w)=p$. Let $i=1,2,\ldots,n$ be such that $\sir_i(\mu)>\mu$, where $\si_i$ are
reflections corresponding to simple roots $\al_i$. Then $l(\si_i w)<l(w)$
(Lemma 3.5) and
dimHom$_g\left(M(\mu),M(\sir_i(\mu))\right)\neq 0$ (Proposition 3.7).
Consider the $sl_2$
subalgebra $g_i$ corresponding to the simple root $\al_i$,
$i=1,2,\ldots,n$. $M(\sir_i(\mu))$
is the so-called completion of $M(\mu)$ w.r.t $g_i$ and is unique
(\cite{Enr}, Proposition 3.6, and \cite{RW1}, Proposition 8.11). Then,
dimHom$_g\left(M(\mu),M(\la)\right)=\mbox{dimHom}_g\left(M(\sir_i(\mu))
,M(\la)\right)$. By the induction hypothesis
$\mbox{dimHom}_g\left(M(\sir_i(\mu))
,M(\la)\right)=1$. This gives the theorem in the case
$k>-\cg $.\vs{2mm}\\
$k<-\cg $: We will prove the theorem using essentially the original
argument of Verma \cite{V}, Theorem 2.
 By Proposition 3.4 it is sufficient to prove that
$\mbox{dimHom}_g\left(M(\la),M(\mu)\right)$ $\leq 1$, where
$\mu=\wro(\la),\ \la\in\Ga_w^-$. As $M(\la)$ is irreducible, we can count the
number of states in $M(\mu)$ and $M(\la)$ to establish that if
dimHom$_g\left(M(\la),M(\mu)\right)\geq 2$, then
$$P(\eta)=\mbox{dim}M_\eta(\mu)\geq
2\mbox{ dim}M_{\eta+\la-\mu}(\la)=2P(\eta+\la-\mu).$$
This is, however, a contradiction \cite{V}, Lemma 3, as can be seen by
considering large $\eta$.$\ \Box$\vs{5mm}\\
As Theorem 3.10 shows that an element of
$\mbox{Hom}_g\left(M(\mu),M(\nu)\right)$ is
either zero or unique (up to a multiplicative constant), we write
$M(\mu)\subset M(\nu)$ whenever
$\mbox{Hom}_g$ $\left(M(\mu),M(\nu)\right) \neq 0$. We next generalize a result
established for finite dimensional Lie algebras \cite{BGG2} and for
$k>-c_{\bar{g}}$
in \cite{RW1}.\vs{5mm}\\
{\sc Theorem 3.11}. Let $\mu,\nu\in\Ga_w$. Then there exist $w,w'\in W$ and
$\la+\rho/2,\la'+\rho/2\in\Ga_w^+\ (k>-\cg )$, or
$\la,\la'\in\Ga^-_w\ (k<-\cg )$ such that $\mu=\wro(\la')$  and
$\nu=w^{\prime\rho}(\la)$. \vs{2mm}\\
For $k>-\cg $ we have:\\
(i)\hs{3mm} $M(\mu)\subset M(\nu)\IFF w\prec w',\ \la=\la'\IFF
\left(M(\nu):L(\mu)\right)\neq 0$\vs{2mm}\\
(ii)\hs{3mm} If $M(\mu)\subset M(\nu)$, $\mu\neq \nu$, then there are
$\mu=\mu_0,\mu_1,\ldots,\mu_{n}=\nu$ such that $\mu_{i+1}=\wro_i(\la)$,
$i=0,1,\ldots n-1$ with
$l(w_{i+1})=l(w_{i})-1$, $w_0=w,w_n=w'$ and
$$M(\mu_0)\subset M(\mu_1)\subset M(\mu_2) \subset \ldots\subset
M(\mu_{n}).\vs{3mm}$$
For $k<-\cg $ we have:\\
(iii)\hs{3mm} $M(\mu)\subset M(\nu)\IFF w'\prec w,\ \la=\la'\IFF
\left(M(\nu):L(\mu)\right)\neq 0$\vs{2mm}\\
(iv)\hs{3mm} If $M(\mu)\subset M(\nu)$, $\mu\neq \nu$, then there are
$\mu=\mu_0,\mu_1,\ldots,\mu_{n}=\nu$ such that $\mu_{i+1}=\wro_i(\la)$,
$i=0,1,\ldots n-1$ with
$l(w_{i+1})=l(w_{i})+1$, $w_0=w,w_n=w'$ and
$$M(\mu_0)\subset M(\mu_1)\subset M(\mu_2) \subset \ldots\subset M(\mu_{n}).$$
{\sc Proof.}(cf \cite{Dix} and \cite{RW1})  Consider $k<-\cg $. The
existence of $w,w'$
follows from Lemma 3.2.
 By Theorem 3.1
and Corollary 3.9 we have $M(\mu)\subset
M(\nu)\IFF\left(M(\nu):L(\mu)\right)\neq 0$.
Assume$M(\mu)\subset M(\nu)$. By Corollary 3.9 there exist
$\ga_1,\ldots,\ga_{n}\in\De^+$
such that
$$\mu=w^{\rho}(\la)<\sir_{\ga_1}w^{\rho}(\la)<\ldots<\sir_{\ga_{n}}\sir_{\ga
_{n-1}}\ldots
\sir_{\ga_1}w^{\rho}(\la)=w^{\prime\rho}(\la').$$
Then $\la=\la'$ (Lemma 3.2) and by Lemma 3.5 we have
$l(w)<l(\si_{\ga_1}w)<\ldots<l(w')$.
Hence, $w\prec w'$ (Lemma 3.7).\vs{2mm}

We now assume $w'\prec w,\ \la=\la'$. Then there exist
$\ga_1,\ldots,\ga_n\in\De^+$ such that
$$w=w_0\stackrel{\ga_1}{\longrightarrow}w_1\stackrel{\ga_2}{\longrightarrow}
w_2\cdots
\stackrel{\ga_{n-1}}\longrightarrow
w_{n-1}\stackrel{\ga_n}{\longrightarrow}w_n=w'.$$
By Lemma 3.5 we have
$\mu=\wro_0(\la)\leq\wro_1(\la)\leq\ldots\leq\wro_n(\la)=\nu$ and, hence,
$$M(\wro_0(\la))\subset M(\wro_1(\la))\subset\ldots\subset M(\wro_n(\la))$$
(Proposition 3.8). This proves (iii) and (iv). The cases (i) and (ii) are proved
analagously. $\Box$\vs{5mm}

It is convinient to introduce the concept of {\it length} of a weight.
Let $\mu\in\Ga_w$. Then we define the length $l(\mu)$ as the smallest integer
$l(w)$ such that $\mu=\wro(\la)$, $w\in W$, $\la+\rho/2\in\Ga_w^+$ or
$\la\in\Ga_w^-$.
We now prove some useful results involving this concept. First we have a
result similar
to Lemma 3.5.\vs{5mm}\\
{\sc Lemma 3.12.} Let $\la+\rho/2\in\Ga_w^+\ (k>-\cg )$ or $\la\in\Ga_w^-\
(k<-\cg )$, $w\in W$ and $\al\in\De^+\cap \Delta^R$.
The following conditions are equivalent\vs{2mm}\\
\hs{5mm}(i)\hs{2mm} $\sir_\al\wro(\la)<\wro(\la)$\\
\hs{5mm}(ii)\hs{2mm} $l\left(\sir_\al\wro (\la)\right)>
l\left(\wro (\la)\right)$ for
$k>-\cg $, or $l\left(\sir_\al\wro (\la)\right)<l\left(\wro (\la)\right)$ for
$k<-\cg $.\vs{5mm}\\
{\sc Proof.}
We prove (i)$\Longrightarrow$ (ii) for the case $k>-\cg $. Let
$w^{\prime\rho}(\la)=\sir_\al
\wro(\la)$ with $l(w')=l(\sir_\al \wro(\la))$. We have
$$\sir_\al w^{\prime\rho}(\la)=\wro(\la)>\sir_\al
\wro(\la)=w^{\prime\rho}(\la).$$
and, thus, $l(w)=
l(\si_\al w')<l(w')$ (Lemma 3.5). By definition, $l(w)\geq l(\wro(\la))$
and, hence,
$$l(\wro(\la))<l(w')=l(\sir_\al \wro(\la)).$$
The case $k<-\cg $ is proved analogously.

We now prove (ii) $\Longrightarrow$ (i) for $k>-\cg $. We
have  $$\sir_\al\wro(\la)=\wro(\la)-n\al.$$
Here $n={(2\wro(\la)+\rho)\cdot\al\over\al^2}\in{\cal Z}$. $n=0$ implies
$\sir_\al\wro(\la)=\wro(\la)$ and thus
$l(\sir_\al\wro(\la))=l(\wro(\la))$. This contradicts (ii) and, therefore, we
have $n\neq 0$. If $n<0$ then $\sir_\al\wro(\la)>\wro(\la)$. By the implication
(i) $\Longrightarrow$ (ii), we again contradict (ii). Hence, $n=1,2,\ldots$ and
(i) follows. The proof for $k<-\cg $ is analogous. $\Box$\vs{5mm}\\
We may easily generalize \cite{Dix}, Proposition 7.6.8, to obtain:\vs{5mm}\\
{\sc Lemma 3.13}. Let $\la+\rho/2\in\Ga_w^+\ (k>-\cg )$ or $\la\in\Ga_w^-\
(k<-\cg )$ and $w=\si_{\al_n}\ldots\si_{\al_1}$ be a reduced decomposition
of $w\in W$,
where $\al_1,
\ldots,\al_n\in\De^s$. Let
$\la_0=\la$, $\la_1=\sir_{\al_1}(\la_0)$, $\la_2=\sir_{\al_2}(\la_1),$
$\ldots,\la_n=\sir_{\al_n}
(\la_{n-1})$. Then for $k>-\cg $
$$\la_0\geq\la_1\geq\ldots\geq\la_n \mbox{ and }\al_{i+1}\cdot(\la_i+\rho/2)
\in\{0,1,2\ldots\}$$
and for $k<-\cg $
$$\la_0\leq\la_1\leq\ldots\leq\la_n \mbox{ and }\al_{i+1}\cdot(\la_i+\rho/2)\in
\{0,-1,-2\ldots\}$$
{\sc Lemma 3.14}. Let $\la+\rho/2\in\Ga_w^+\ (k>-\cg )$ or $\la\in\Ga_w^-\
(k<-\cg )$. Let $\mu\in\Ga_w$ with $\mu=\wro(\la)$, $w\in W$. If
$l(\mu)=l(w)$, then
$w,\la,\mu$
satisfy (**) in Lemma 3.2 with $l(\mu)=n$. In addition, this is the minimal
integer $n$ for
which (**) is satisfied.
\vs{5mm}\\
{\sc Proof.} Consider $k<-\cg $. Let $w=\si_{\al_n}\ldots\si_{\al_1}$ with
$l(w)=l(\mu)$.
By Lemma 3.13 we have a sequence  $$\la_0\leq\la_1\leq\ldots\leq\la_n \mbox{ and
}\al_{i+1}\cdot(\la_i+\rho/2)\in\{0,-1,-2,\ldots\}.$$ Assume $\la_i=\la_{i+1}$
for some $i\in\{0,1,2\ldots,n\}$. Then clearly
$w'=\si_{\al_n}\ldots\si_{\al_{i+1}}\si_{\al_{i-1}}\ldots \si_{\al_1}$
satisfies
$\mu=w^{\prime\rho}(\la)$ and $l(w')<l(w)$. This contradicts the assumption
$l(\mu)=l(w)$.
The last assertion follows by the definition of $l(\mu)$. $k>-\cg $ is
proved analagously.
$\Box$\vs{5mm}\\
{\sc Proposition 3.15}. Let $M(\mu)\subset M(\nu)$, where $\mu,\nu\in\Ga_w$.
Then $l(\mu)-l(\nu)=n$ for $k>-\cg $, or $l(\nu)-l(\mu)=n$ for $k<-\cg $, if and
only if $n$ is the largest integer for which $M(\mu)\subset
M^{(n)}(\nu)$.\vs{5mm}\\
{\sc Proof.} Consider $k<-\cg $. By Proposition 3.4 and the hereditary nature of
Jantzen's filtration it is sufficient to prove the proposition for $l(\mu)=0$
i.e. for $\mu\in\Ga_w^-$ and some given $M(\nu)$. We prove the "only if"
case by
induction on $l(\nu)$. For
$l(\nu)=0$ the proposition is trivial. Assume it to be true for $l(\nu)\leq
p-1$ and
consider $l(\nu)=p$. As $p\geq 1$ there must exist $\al\in\De^s$ such that
$\nu'=\sir_\al(\nu)<\nu$. Then $M(\nu')\subset M(\nu)$ (Proposition 3.8) and $l(\nu')<l(\nu)$
(Lemma 3.12). If $l(\nu')<p-1$ then $l(\nu)<p$, which is a contradiction. Hence,
$l(\nu')=p-1$. In addition, $M(\nu')\subset M^{(1)}(\nu)$ and
$M(\nu')\not\subset
M^{(2)}(\nu)$. This follows by an explicit construction of the highest
weight vector that
generates $M(\nu')$ (cf. the proof of Lemma 3.3). We now use the induction
hypothesis
on $M(\nu')$ together with the hereditary nature of Jantzen's filtration to
conclude that the proposition holds for $l(\nu)=p$.

We prove the "if" case. Consider $M(\mu)\subset M^{(p)}(\nu)$,
$\mu\in\Ga_w^-$ and use
induction on $p$. The case $p=0$ is trivial. Assume the assertion to be
true for $0\leq
p\leq n-1$ and consider $p=n\geq 1$. As $p\geq 1$ there must exist
$\al\in\De^s$ such that
$\nu'=\sir_\al(\nu)<\nu$ and $M(\nu')\subset M(\nu)$ (Proposition 3.8)
with $l(\nu')<l(\nu)$ (Lemma 3.12). By explicit construction of the highest
weight
vector one checks that $M(\nu')\subset M^{(1)}(\nu)$ and $M(\nu')\not\subset
M^{(2)}(\nu)$. Then the hereditary nature of Jantzen's filtration implies
$M(\mu)\subset
M^{(n-1)}(\nu\pri)$, which by the induction hypothesis yields
$l(\nu\pri)=n-1$. Then
$\nu\pri=\sir_\al(\nu)$ implies $l(\nu)=n$, which concludes the proof.
The case $k>-\cg $ is proved in a completely analogous fashion. $\Box$\vs{5mm}\\
{\sc Lemma 3.16.} (\cite{Dix}, Lemma 7.7.6; \cite{RW1}, Lemma 8.6). Let
$w_1,w_2\in W$. The
number of elements
$w\in W$ such that $w_1\leftarrow w\leftarrow w_2$ is 0 or 2. \vs{5mm}\\
{\sc Lemma 3.17.} (cf. \cite{Dix}, Lemma 7.7.7 (iii) and  \cite{RW1}, Lemma
8.15 (iii)) Let
$M(\mu_1)$ and
$M(\mu_2)$ be Verma modules with  highest weights $\mu_1$ and $\mu_2$,
respectively. Let
$\mu_1+\rho/2$ and $\mu_2+\rho/2$ be regular. If
$l(\mu_1)=l(\mu_2)+2$ $(k>\cg)$ or
$l(\mu_1)=l(\mu_2)-2$ $(k<\cg)$, then the number of $\mu$
such that $M(\mu_1)\subset M(\mu)\subset M(\mu_2)$, $M(\mu_1)\neq
M(\mu)\neq M(\mu_2)$
is either 0 or 2.
\vs{5mm}\\ {\sc Proof.} Consider $k>-\cg$. The definition of $l(\mu_1)$ and
$l(\mu_2)$
implies together with Lemma 3.2 that there exists $w_1,w_2\in W$ such that
$\mu_1=\wro_1(\la)$, $\mu_2=\wro_2(\la)$, $\la+\rho/2\in \Ga_w^+$ and
$l(w_1)=l(w_2)+2$. In addition, $\mu_1+\rho/2$ and $\mu_2+\rho/2$ regular
imply that
$\la\in\Ga_w^+$. Then the number of
$w\in W$ such that
$M(\wro_1(\la))\subset M(\wro(\la))\subset M(\wro_2(\la))$ and
$M(\wro_1(\la))\neq
M(\wro(\la))\neq M(\wro_2(\la)) $ is 0 or 2, as can be seen from combining
Lemma 3.16 and
Theorem 3.11. This proves the assertion of the lemma for $k>-\cg$. The case
$k<-\cg$ is
proved analogously. $\Box$\vs{5mm}\\

\section{ The BRST formalism}
\setcounter{equation}{0}
Define the algebra $g'=g_k\oplus g_{-k-2c_{\bar{g}}}$, where
$c_{\bar{g}}$ is the quadratic
Casimir of the adjoint representation. This algebra is invariant under the
exchange
$k\rightarrow -k-2c_{\bar{g}}$ and, consequently, we may restrict to
$k>c_g$. The singular
case $k=-c_{\bar{g}}$ will not be treated here. We will denote
$g_{-k-2c_{\bar{g}}}$ by $\tg$ and
the Verma module over $\tg$ will be denoted $\widetilde{M}(\ti{\la})$,
 where $\ti{\la}$ is its
highest weight. Let $B_{n'_+}$, $B_{n'_-}$, $B_{h'}$, $B_{\ti{g}}$ and
$B_{g'}$ be bases
of $n'_+$, $n'_-$, $h'$, $\ti{g}$ and $g'$, respectively. The generators
$\ti{e}_\al,\
\ti{f}_\al \mbox{ and }\ti{h}_i$  is a realization
of $B_{\tg}$ and $e'_\al,\ f'_\al\mbox{ and }h'_i$ a
realization of $B_{g'}$. Define
$M'_{\la\ti{\la}}=M(\la)\otimes\tM(\ti{\la})$ and similarly
$L'_{\la\ti{\la}}=L(\la)\otimes\ti{L}(\ti{\la})$. $\pi_{L'}$ denotes the
projection
$M'\longrightarrow L'$.

We define the anticommuting ghost and antighost operators $c(x)$ and $b(x)$,
respectively, where $x\in B_{g'}$, with the following properties
\be
&(i)&\{c(x),b(y)\}=\delta_{x^\dagger,y}\\
&(ii)& c^\dagger(x)=c(x^\dagger),\ b^\dagger(x)=b(x^\dagger)\\
&(iii)& b(a_1x+a_2y)=a_1b(x)+a_2b(y)\hspace{5mm} a_1,a_2\in{\cal C}.\ee
Here $\delta_{x,y}=1$ if $x=y$ and $0$ otherwise. Introduce a normalordering
\be
:c(x)b(y):\ =\left\{\begin{array}{ll}\  c(x)b(y) &\mbox{ if either } x\in B_{n'_-}
\mbox{ or } y\in
B_{n'_+}\\ -b(y)c(x) &\mbox{ if  either } x\in B_{n'_+} \mbox{ or } y\in
B_{n'_-}\\
\frac{1}{2}(c(x)b(y)-b(y)c(x)) &\mbox{ otherwise}\end{array}\right..
\ee
Define the BRST operator
\be
d=\sum_{x\in B_{g'}}c(x^\dagger)x+\sum_{x\in
B_{h'}}c(x^\dagger)\rho(x)-\frac{1}{2}
\sum_{x,y\in B_{g'}}:b([x,y])c(x^\dagger)c(y^\dagger):,\ee
where $\rho(x)$ is the component of $\rho$ corresponding to the element
$x\in B_{h'}$.
 The BRST operator has the
following two fundamental properties: $d^2=0$ and $d^\dagger=d$. The first
property implies that $x^{tot}=\{d,b(x)\}=x+\rho+\sum_y b([x,y])c(y^\dagger)$
generates an algebra $g_0$ which is centerless.

Define a ghost module ${\cal F}^{gh}$. It is generated by the ghost operators
acting on a vacuum vector $v_0^{gh}$ satisfying
\be
c(x)v_0^{gh}=b(y)v_0^{gh}=0
\mbox{ for } x\in B_{n'_+} \mbox{ and }y\in B_{n'_+}\cup B_{h'} .
\ee
We also define a restricted module $\hat{{\cal F}}^{gh}=\{v^{gh}\in{\cal
F}^{gh}\
|\ b(x)v^{gh}=0 \mbox{ for } x\in B_{h'}\}$. The dual ${\cal F}^{\ast
gh}$ of ${\cal F}^{gh}$ has a vacuum vector $v_0^{\ast gh}$ satisfying
\be
c^\dagger (x)v_0^{\ast gh}=b^\dagger (y)v_0^{\ast gh}=0 \mbox{ for } x\in
B_{n'_+}\cup
B_{h'}
\mbox{ and } y\in B_{n'_+} .
\ee
The restricted dual is $\hat{{\cal F}}^{\ast gh}=\{v^{\ast gh}\in{\cal F}^{\ast
gh}\ |\ c(x)v^{\ast gh}=0\mbox{ for } x\in B_{h'}\}$.
Define a Hermitean form for the ghost sector by
\be
\lan v_0^{\ast gh}|v_0^{gh}\ran=1\nonu
\lan v^{\ast gh}|uv^{gh}\ran=\lan u^\dagger v^{\ast gh}|v^{gh}\ran,\ee
for a polynomial $u$ in the ghost operators and $v^{gh}\in {\cal F}^{gh},
\ v^{\ast gh}\in{\cal
F}^{\ast gh}$. If $v^{gh}=uv_0^{gh}$
then we denote by $v^{\ast gh}$ the vector $u^\dagger v_0^{\ast gh}$.

The ghost number $N^{gh}$ of any vector $v^{gh}\in {\cal F}^{gh}$
is defined by $N^{gh}(v_0^{gh})=0$ and $ N^{gh}(c(x)v)=N^{gh}(v)+1,\
N^{gh}(b(x)v)=N^{gh}(v)-1$. The ghost numbers of vectors in the
dual module is
similarly defined with $N^{gh}(v_0^{\ast gh})=0$. It is easily seen that
$\lan u^\ast|v\ran=0$ if $N^{gh}(u^\ast)+N^{gh}(v) \neq  0$. Let $C(g',V)$
be the
complex $V\otimes{\cal F}^{gh}$ for a $g'$-module $V$. We define the relative
subcomplex $\hat{C}(g',V)=\{\omega\in C(g',V)\ |\ b(x)\omega=0,\
x^{tot}\omega=0$ for $x\in B_{h'}\}$ and $\hat{C}(g',V^\ast)$ is the dual
complex.
If $\om=v\otimes v^{gh}$ for $v\in V$, $v^{gh}\in {\cal F}^{gh}$, then we denote
by $\om^\ast$ the vector $v\otimes v^{\ast gh}$.
We decompose $d$ as
\be
d=\hat{d}+\sum_{x\in B_{h'}}(x^{tot}c(x)+{\cal M}(x)b(x)).
\ee
We have $d\omega=\hat{d}\omega$ for $\omega\in\hat{C}(g',V)$ and consequently
on the relative subcomplex we may analyze the cohomology of $\hat{d}$ in place
of $d$. The cohomology associated with $\hd$, the semi-infinite or BRST relative
cohomology is sometimes denoted by $\hat{H}^{\infty/2+p}(g',V)$ to
distinguish it from
the conventional Lie algebra cohomology. We will, however, here for
simplicity write
$\hat{H}^p(g',V)$, where p refers to elements $\om$ with $N_{gh}(\om)=p$.
Our primary
interest here will be for $V=L'_{\la,\ti{\la}}$. But in order to gain
knowledge of this
case we will also study $V=M'_{\la,\ti{\la}}$ and its submodules.

It will be convinient to make a classification of vectors in the complex
$C(g',V)$
using the BRST operator. A central result due to Kugo and Ojima \cite{KO}
states the following.\vs{3mm}\\
{\sc Theorem 4.1.} Let $V$ be an irreducible module. Then a basis of
$C(g',V)$ may be
chosen so that for an element $\om$ in this basis one of the following will
be true.
\\ {\em (i) Singlet case:} $\om\in H^\ast(g',V)$ and $\lan\om^\ast|\om\ran
\neq 0$, $N^{gh}(\om)=0$.\\
{\em (ii) Singlet pair case:} $\om\in H^\ast(g',V)$ and there exists an element
$\psi\neq\om$
such that $\psi\in H^\ast(g',V)$, $\lan\psi^\ast|\om\ran\neq 0$ and
$N^{gh}(\psi)=-N^{gh}(\om)$.\\
{\em (iii) Quartet case:} $\om\not\in H^\ast(g',V)$. There will exist
four elements  $\om_1,\om_2,
\psi_1,\psi_2\in C(g',V)$, where
either $\om=\om_1 \mbox{ or }\om=\om_2$, such that
$\lan\psi_1^\ast|\om_1\ran\neq 0$ and $\lan\psi_2^\ast|\om_2\ran\neq 0$,
$\om_2=d\om_1$ and $\psi_1=d\psi_2$ and $N^{gh}(\om_1)=N^{gh}(\om_2)-1
=-N^{gh}(\psi_1)=-N^{gh}(\psi_2)-1$.\vs{3mm}\\
There will exist an analagous classification on the relative subcomplex
$\hat{C}(g',V)$ w.r.t. $\hat{d}$. In this classification all non-trivial
elements
in the cohomology will be singlets or singlet pairs. It should be remarked
that the condition of irreducibility is essential for the theorem. In the
following section, we will find that for $V$ being a reducible Verma module
the above classification does not hold. In particular, the non-trivial
elements of the cohomology for non-zero ghost numbers will for this case
not be members of  singlet pairs. Define Jantzen's filtration for
$\xi\in\hat{C}(g',M')$ as follows. Let $\la_\ep=\la+\ep z$ and $\ti{\la}_\ep=
\ti{\la}-\ep z$. Then $M^{\prime (n)}(\la_\ep)=\{v'\in
M(\la_\ep)\otimes\widetilde{M}
(\ti{\la}_\ep)|\ \lan w^{\prime\ast}|v'\ran\mbox{ is divisable by }\ep^n
\mbox{ for any } w^{\prime \ast}\in M^\ast(\la_\ep)\otimes\widetilde{M}^\ast
(\ti{\la}_\ep)\}$. We denote by $\xi_\ep$ the vector $v_\ep\otimes
\ti{v}_\ep\otimes v^{gh}$.
An element $\xi_\ep$ is always assumed to be finite as $\ep\rightarrow 0$.
We denote by
$f(\ep)\sim\ep^n$ the leading order of a function $f(\ep)$ in the limit
$\ep\rightarrow 0$.
Note that our definition of Jantzen's filtration for $g'$ implies that
$\la_\ep+\ti{\la}_\ep$
is independent of $\ep$. This is required if the cohomology should
have at least one
non-trivial element for $\ep\neq 0$, namely the vacuum solution
$\xi_{0\ep}=v_{0\ep}\otimes\ti{v}_{0\ep}\otimes v_0^{gh}$.

 In the next section the following result will
be needed. \vs{5mm}\\
{\sc Lemma 4.2}. Let $\xi_{1\ep},\xi_{2\ep}\in\hat{C}(g',M'_\ep)$ be
non-zero for $\ep=0$ and $\hd\xi_{2\ep}=g(\ep)\xi_{1\ep}$, where $g(\ep)\sim 1
\mbox{ or } \ep$. Let $n_1$ and $n_2$ be the largest integers for which
$\xi_{i}\in\hat{C}(g',M^{\prime(n_i)})$, $i=1,2$. Then there exist
$\ze_{1\ep},\ze_{2\ep}\in\hat{C}(g',M'_\ep)$ which are non-zero for
$\ep=0$ and satisfy\vs{1mm}\\
  \hs{5mm}(i) $\lan\ze^\ast_{i\ep}|\
\xi_{j\ep}\ran\sim\ep^{n_i}\del_{i,j}\neq 0$ for $\ep\neq 0$,
$i=j=1,2$.\vs{1mm}\\
 \hs{4mm}(ii) $\hd\ze_{1\ep}=f(\ep)\ze_{2\ep}$, where $f(\ep)\sim 1\mbox{ or
}\sim \ep$.\vs{1mm}\\
\hs{3mm}(iii) $n_1,\ n_2$ are the largest integers
for which $\ze_i\in\hat{C}(g',M^{\prime(n_i)})$.\vs{1mm}\\
 In addition, for
$g(\ep)\sim 1$:  $n_1=n_2$ if and only if $f(\ep)\sim 1$, $n_1=n_2+1$ if and
only if $f(\ep)\sim\ep$. For $g(\ep)\sim\ep$ we have: $n_1=n_2-1$ if and only if
 $f(\ep)\sim 1$,
$n_1=n_2$ if and only if $f(\ep)\sim\ep$. \vs{3mm}\\
{\sc Proof.} Since
$\xi_{1,\ep}\in\hat{C}(g',M^{\prime(n_1)}_\ep)$ for a largest integer $n_1$ and
$M^{\prime(n_i)}_\ep$ is irreducible for $0<|\ep|\ll 1$ there must exist
one vector
$\ze_{1\ep}\in\hat{C}(g',M^{\prime(n_1)}_\ep)$ with $\lan\xi^\ast_{1\ep}|
\ze_{1\ep}\ran\sim\ep^{n_1}$. Then
\be
g(\ep)\lan\ze^\ast_{1\ep}| \xi_{1\ep}\ran=\lan\ze^\ast_{1\ep}|\hd\xi_{2\ep}\ran=
\lan\hd\ze^\ast_{1\ep}| \xi_{2\ep}\ran\label{eqi}
\ee
implies $\hd\ze_{1\ep}=f(\ep)\ze_{2\ep}$, for some vector $\ze_{2\ep}$
satisfying $\lan\ze^\ast_{2\ep}| \xi_{2\ep}\ran\neq 0$ and which is non-zero
for $\ep=0$. In addition, $f(\ep)$ is a non-singular function of
$\ep$. From the fact that $\hd$ is linear in the generators of $g'$ we can can
conclude that $f(\ep)\sim 1\mbox{ or }\ep$.
 Pick a basis of $\hat{C}(g',M'_\ep)$ such that $\xi_1,\ \xi_2$ are two of its
elements. Denote the elements of the basis by $\xi_i$, $i=1,2,3,\ldots$ .
Similarly we pick a basis of
$\hat{C}(g',M^{\prime\ast}_\ep)$, $\ze^{\ast}_{i\ep}$, $i=1,2,3, \ldots$ .
We choose it such that $\lan\ze_{i\ep}^{\ast}|\xi_{j\ep}\ran$ is non-zero
only for $i=j$. Now since $\lan\ze_{i\ep}^\ast|\xi_{2\ep}\ran=0$ for
$i\neq 2$ and $\xi_{2\ep}\in\hat{C}(g',M^{\prime(n_2)}_\ep)$ we must have
$\lan\ze_{2\ep}^\ast|\xi_{2\ep}\ran\sim\ep^{n_2}$. This in turn implies,
using $\lan\ze_{2\ep}^\ast|\xi_{j\ep}\ran=0$ for $j\neq 2$ and the
definition of Jantzen's filtration, that $\ze_{2\ep}\in
\hat{C}(g',M^{\prime(n_2)}_\ep)$. We now conclude from $\lan\xi_{1\ep}^\ast|
\ze_{1\ep}\ran\sim\ep^{n_1}$, $\lan\xi_{2\ep}^\ast|
\ze_{2\ep}\ran\sim\ep^{n_2}$, eq.(\ref{eqi}) and $f(\ep)\sim
1\mbox{ or }\ep$ that for
$g(\ep)\sim1$ we have $\ep^{n_1}\sim\ep^{n_2}f(\ep)\sim\ep^{n_2}\mbox{ or }
\ep^{n_2+1}$, while for $g(\ep)\sim\ep$ we have
$\ep^{n_1}\sim\ep^{n_2-1}f(\ep)\sim\ep^{n_2-1}\mbox{ or } \ep^{n_2}$. \
$\Box$\vs{3mm}

A standard tool in the analysis of the cohomology is a contracting homotopy
operator. Let $\om_0$ be a vacuum vector of $\hat{C}(g',M')$ i.e.
$\om_0=v'_0\otimes v_0^{gh}$, where $v'_0=v_0\otimes\ti{v_0}$ and $v_0,
\ti{v}_0$
are primary highest weight vectors of $M$ and $\ti{M}$, respectively.
Consider an element $\om\in \hat{C}(g,M')$ of the form $\om=v'\otimes v^{gh}$
with $v'=uv_0\otimes \ti{u}v_0, u\in{\cal U}(n_-) ,\ \ti{u}\in {\cal
U}(\ti{n}_-)$ and
$N^{gh}(v^{gh})=n$. We write $u=u_m+u_{m-1}+\ldots +u_0$, where
$u_i\in{\cal U}(n_-)$ is a monomial of order $i$. Introduce a gradation
$N_{gr}$. We define $N_{gr}(\om_0)=0$. Furthermore, $N_{gr}(\om)=m-n$.
We will get a filtration $\hat{C}(g',M')=\bigoplus_{N_{gr}}\hat{C}(g',M')
_{N_{gr}}$. We now decompose $\hd$ as $\hd=d_0+d_{-1}$, where $d_0=
\sum_{\al\in \De^+}c(e'_\al)\ f_\al$. We have $d\om=d_0\om+
(\mbox{lower order terms})$. Let $\om_{p,q}\in\hat{C}(g',M')_
{p+q-r}$ be of the form
\be
\om_{p,q}=f_{\al_1}\ldots f_{\al_p}v_0\otimes \ti{v}\otimes
b(f'_{\beta_1})\ldots b(f'_{\beta_q})c(f'_{\ga_1})\ldots c(f'_{\ga_r})v_0^{gh},
\ee
where $\al,\beta,\ga\in\De^+$. The homotopy operator
$\ka_0$ is now defined by
\be
&\ka_0\om_{p,q}={1\over p+q}&\sum_{i=1}^p f_{\al_1}\ldots
\widehat{f_{\al_i}}\ldots f_{\al_p}v_0\otimes \ti{v}\otimes
b(f'_{\al_i})b(f'_{\beta_1})\ldots b(f'_{\beta_q})\nonu
&& c(f'_{\ga_1})\ldots c(f'_{\ga_r})v_0^{gh}\hs{20mm}p\neq 0\nonu
&\ka_0\om_{0,q}=0,&
\ee
where capped factors are omitted. It is now straightforward to verify
\be
(d_0\ka_0+\ka_0d_0)\om_{p,q}=(1-\delta_{p+q,0})\om_{p,q}+
(\mbox{lower order terms}) \label{kappad}.
\ee
 One may also define a gradation $\ti{N}_{gr}$ using the elements of
${\cal U}(\ti{n}_-)$ in place of ${\cal U}(n_-)$.
We then have a corresponding decomposition $\hd=\ti{d}_0+\ti{d}_{-1}$
with $\ti{d}_0=\sum_{\al\in \De^+}c(e'_\al)\ \ti{f}_\al$ and a
homotopy operator $\ti{\ka}_0$.

\section{The BRST cohomology}
\setcounter{equation}{0}

We will now in detail study the semi-infinite relative cohomology
associated with the BRST
operator. The notation follows that of previous sections.
$\om_\ep,\xi_\ep,\ldots$ always
denote elements of $\hat{C}(g',\ldots)$ that are finite in the limit
$\ep\rightarrow 0$. Our starting point is  Proposition 5.1 concerning the
cohomology of  Verma modules. This proposition was to our knowledge first given
in \cite{LZ}, Proposition 2.29.\vs{5mm}\\
{\sc Proposition 5.1}. Let
$M'$ be a highest weight Verma module over  $g'$. Then $\hat{H}^p (g',M')=0$ for
$p<0$.\vs{3mm}\\   {\sc Proof.}(\cite{HR}) Let $\om\in\hat{H}^p (g',M')$
and have a highest
order term  $\om_n$ in the gradation $N_{gr}$. Then $0=\hd\om=d_0\om_n+
(\mbox{lower order terms})$ and hence $d_0\om_n=0$ to leading order.  Using
eq.(\ref{kappad}) we conclude that $\om_n=d_0(\ka_0\om_n)+(\mbox{lower order
terms})$ and as a consequence of this, $\om=\hd(\ka_0\om_n)+(\mbox{lower order
terms})$. Thus $\om$  is a trivial element of $\hat{H}^p (g',M')$ to highest
order. This may  be iterated to lower orders and we find that $\om\in\hat{H}^p
(g',M')$  will be non-trivial only for $N_{gr}(\om)\leq 0$, which is
impossible if
$N^{gh}(\om)<0$. $\Box$ \vs{5mm}
\\
{\sc Corollary 5.2}. (\cite {HR}) Let $M'$ in Proposition 5.1 be irreducible.
Then $\hat{H}^p (g',M')=0$ for $p\neq 0$. Furthermore,
$\om\in\hat{H}^0 (g',M')$
is the element $\om=v_0\otimes \ti{v}_0\otimes v_0^{gh}$, where $v_0$ and
$\ti{v}_0$
are primary highest weight vectors of weights $\la$ and $\ti{\la}$,
respectively,
satisfying $\la+\ti{\la}+\rho=0$.  \vs{5mm}
\\
{\sc Corollary 5.3}. Let $L'$ be the irreducible $g'$-module of $M'$. Let
$\om\in\CM$ be such that $0\neq\pil(\om)\in \hat{H}^p (g',L'),\ p<0$. Then
\be\hd\om=\nu,\ee
where $\nu\in \hat{H}^{p+1} (g',M^{\prime (1)})$ and is non-zero.\vs{3mm}\\
{\sc Proof.}(\cite{HR}) Assume first $\hd\om=0$ in $\CM$. Then Proposition 5.1
implies $\om=\hd\eta,\ \eta\in\hat{C}(g',M')$. Since $\om\in\hat{C}
(g',M'/M^{\prime (1)})$ we must have $\eta\in\hat{C}(g',M'/M^{\prime (1)})$,
which implies that $\om$ is cohomologically trivial. Therefore, $\hd\om=\nu
\neq 0$ and so $\hd\nu=0$. If $\nu\not\in \hat{H}^{p+1} (g',M^{\prime (1)})$,
then $\nu=\hd\nu'$ for some $\nu'\in\hat{C}(g',M^{\prime (1)})$ and
$\hd(\om-\nu')=0$. Proposition 5.1 then implies that $\pil(\om)$ is a
trivial element
of $\hat{H}^p (g',L')$.
$\Box$\vs{5mm}\\
The following lemma is partly the converse of Corollary 5.3.\vs{5mm}\\
{\sc Lemma 5.4}. Let $\om\in\CM$,
$\hd\om=\nu$ in $\CM$ with $\nu\in\hat{H}^{p+1} (g',M^{\prime (1)})$ and
$\pil(\om)\neq 0$,
then  $\pil(\om)\in\hat{H}^p (g',L')$.\vs{3mm}\\
{\sc Proof.} Firstly, $\hd\om=\nu$ with $\nu\in\hat{H}^{p+1}
(g',M^{\prime (1)})$ implies
that $\hd\pil(\om)=0$. Secondly, assume
$\pil(\om)$ to be trivial i.e. $\pil(\om)=\hd\pil(\psi),\
\psi\in\hat{C}(g',M')$. Then
$\om=\hd\psi+\nu'$ in $\CM$, with $\nu'\in\CI$, and so $\nu=\hd\om=\hd\nu'$.
This is a contradiction
to the assumption $\nu\in\hat{H}^{p+1} (g',M^{\prime (1)})$.
 $\Box$\vs{5mm}\\
{\sc Lemma 5.5}. dim$\left(\hat{H}^{p+1}
(g',M^{\prime (1)})\right)=\mbox{dim}\left(\hat{H}^p (g',L')\right)$ for $p\leq
-2$.\vs{3mm}\\ {\sc Proof.} Let $\nu\in\hat{H}^{p+1}
(g',M^{\prime (1)})$
with $p\leq -2$, then by Proposition 5.1 $\nu=\hd\om$, $\om\in\CM$ and
$\pil(\om)\neq 0$.
Lemma 5.4
then implies
$\pil(\om)\in\hat{H}^p (g',L')$. We have thus proved
that $\mbox{dim}(\hat{H}^{p+1}(g',M^{\prime (1)}))\leq\mbox{dim}
(\hat{H}^{p}(g',L'))$. We now prove that the dimensionalities
are in fact the same.  Assume two elements $\om_1,\ \om_2\in\CM$ with
$\pil(\om_1),\pil(\om_2)\in\hat{H}^{p}(g',L')$,  corresponding to the same
element $\nu$. By
Corollary 5.3  we have in $\CM$: $\hd\om_1=\nu_1$
and $\hd\om_2=\nu_2$, where $\nu_1=\nu_2$ as elements in
$\hat{H}^{p+1}(g',M^{\prime (1)})$. Subtracting the equations yields
$\hd(\om_1-\om_2)=\nu_1-\nu_2=\hd\nu'$, $\nu'\in\CI$,
which by Proposition 5.1 implies $\om_1-\om_2-\nu'=\hd(\ldots)$.
$\pil(\om_1)$ and $\pil(\om_2)$ are therefore identical elements in
$\hat{H}^p (g',L')$. $\Box$\vs{5mm}\\
The results obtained so far are of importance for negative ghost numbers. We now
turn to results relevant for positive ghost numbers. We will connect the
two cases by the use
of  Jantzen's perturbation, Theorem  4.1 and Lemma 4.2.\vs{5mm}\\
{\sc Lemma 5.6}. Let $\om\in\CM$ with $\pil(\om)\in\hat{H}^p (g',L')$,
satisfying
$\hd\om=\nu$, $\nu\in\CI$. Then: \vs{2mm}\\
\hs{5mm}(i) There exists $\psi\in\CM$ with $\pil(\psi)\in\hat{H}^{-p}
(g',L')$ and
$\lan\psi^\ast |  \om\ran\neq 0$.\vs{1mm}\\
\hs{4mm}(ii) With $\psi$ as in (i): There exists $\chi\in \CI$ of opposite
ghost number of $\nu$,\\
\hs{4mm}satisfying $\hd\chi_\ep=
\ep\psi_\ep$. \vs{1mm}\\
\hs{3mm}(iii) $\chi,\ \nu\not\in\hat{C}(g',M^{\prime (2)})$, where $\chi$
is defined
as in (ii).\vs{1mm}\\
\hs{3mm}(iv) With $\psi$ as in (i): $\hd\psi=0$.\vs{1mm}\\
\hs{4mm}(v) $p\leq 0.$\vs{3mm}
\\
{\sc Proof.} (i) follows directly from Theorem 4.1. (ii) and (iii)
follow from Theorem 4.1 and Lemma 4.2 using $\om=\xi_2,\ \nu=\xi_1$ and
$g(\ep)\sim 1$. (iv) follows by applying $\hd$
to the equation $\hd\chi_\ep=\ep\psi_\ep$, using $\hd^2=0$ and
taking the limit $\ep\rightarrow 0$. Finally (v) may be proved by contradiction.
If $p>0$, then by (iv) and Corollary 5.3 $\psi$ is $\hd-$exact and, hence, so is
$\om$. \ $\Box$\vs{5mm}\\
{\sc Lemma 5.7}. Let
$\psi\in\CM$, $\pil(\psi)\neq 0$, and $\chi\in\hat{C}(g',M^{\prime (1)})$
be such that
$\hd\chi_\ep=\ep\psi_\ep$. Then $\chi\in\hat{C}(g',M^{\prime (1)}
/M^{\prime(2)}),\ \pil(\psi)\in\HL$ and $p\geq 0$. Conversely, let
$\pil(\psi)\in\HL,\ p\geq 1$, then there exists $\chi\in\hat{C}(g',M^{\prime
(1)})$ such that $\hd\chi_\ep=\ep\psi_\ep$.\vs{3mm}\\
{\sc Proof.}  Assume $\hd\chi_\ep=\ep\psi_\ep$. We
apply Lemma 4.2 with $\xi_1=\psi$, $\xi_2=\chi$ and $g(\ep)\sim\ep$. Then
$n_1=0$,
$n_2\geq 1$ and by the lemma there exist two vectors  $\om$ and $\nu$ such that
$\lan \om^{\ast}_\ep|\psi_\ep\ran\sim 1$, $\lan\nu_\ep^\ast|\chi_\ep\ran\sim\ep$
and $\hd\om_\ep=f(\ep)\nu_\ep$,  with $f(\ep)\sim 1\mbox{ or } \ep$. Furthermore
$f(\ep)\sim 1$, since otherwise $n_1=n_2$. This in turn implies $n_2=1$, by
Lemma 4.2
(iii), and $\chi,\nu\in \hat{C}(g',M^{\prime (1)}/M^{\prime (2)})$.
We now show that $\nu$ is not exact in $\hat{C}(g',M^{\prime (1)})$.
Assume the contrary i.e. $\nu=\hd\eta$ with $\eta\in
\hat{C}(g',M^{\prime (1)}/M^{\prime (2)})$. Then
$\nu_\ep=\hd\eta_\ep+h(\ep)\nu'_\ep$, where $\nu'\in\CI$ and $h(\ep)$ is a
polynomial in $\ep$ such that $h(0)=0$. This implies that $\om'_\ep=\om_\ep-
f(\ep)\eta_\ep$
satisfies $\hd\om'_\ep=f(\ep)h(\ep)\nu'_\ep$. Now $\lim_{\ep\rightarrow 0}
\lan\om^{\prime\ast
}_\ep|\psi_\ep\ran\neq 0$ since $\om'$ and $\om$ differ by an element in
$\CI$. This
is a contradiction as can be seen from
$$ \lan\om^{\prime\ast}_\ep
|\psi_\ep\ran=\lan\om^{\prime\ast}_\ep
 | \frac{1}{\ep}\hd\chi_\ep\ran
=\lan\hd\om^{\prime\ast}_\ep | \frac{1}{\ep}\chi_\ep\ran=
\lan f(\ep)h(\ep)\nu^{\prime\ast}_\ep | \frac{1}{\ep}\chi_\ep\ran
\longrightarrow 0\hs{3mm}
\mbox{ for } \ep\rightarrow 0.$$
Thus $\nu\in\hat{H}^{-p+1}(g',M^{\prime (1)})$.
Lemma 5.4 then gives
$\pil(\om)\in\hat{H}^{-p} (g',L')$, which implies $\pil(\psi)\in\HL$.
The condition $p\geq 0$ follows from Corollary 5.3 and the fact that $\hd\psi=0$
in $\CM$.

We now prove the converse statement. Let $\pil(\psi)\in\HL,\ p\geq 1$. Pick
a basis as in
Theorem 4.1 so that $\psi$ is one of its elements and $\om\in\CM$,
$\pil(\om)\in\hat{H}^{-p}
(g',L')$, $\lan\om^\ast|\psi\ran\neq 0$, is another. Corollary 5.3 implies
$\hd\om=\nu$ and
then Lemma 5.6 gives the assertion.
 $\Box$\vs{5mm}\\
{\sc Lemma 5.8} . Let $\psi$ and $\chi\in \CM$ be such
that $N^{gh}(\psi)\geq 1$, $\hd\chi_\ep=\ep\psi_\ep$ and
$\chi\in\hat{C}(g',M^{\prime (1)}/M^{\prime (2)})$.
Then $\pil(\psi)\in\HL$.\vs{3mm}\\
{\sc Proof.} By Lemma 5.7 it is sufficient to prove that $\pil(\psi)\neq 0$.
Assume the contrary i.e. $\psi\in\CI$. Then Lemma 4.2 implies that there
exist two vectors $\om$ and $\nu$ satisfying $\hd\om_\ep=f(\ep)\nu_\ep$ in
$\CM$, where
$f(\ep)\sim \ep$. In addition,  $\psi,\
\om,\ \nu\in\hat{C} (g',M^{\prime (1)}/M^{\prime (2)})$ and
$\lan\om^\ast_\ep|\psi_\ep\ran\sim\ep$, $\lan\nu^\ast_\ep|\chi_\ep\ran\sim\ep$.
Now $N^{gh}(\om)\leq -1$, so that by proposition 5.1, $\om=\hd\om'$ for some
vector $\om'$. We then have  $\om_\ep=\hd\om'_\ep+h(\ep)\nu'_\ep$ for some
vector
$\nu'_\ep$, which is non-singular for $\ep=0$ and $h(\ep)$ is a polynomial of
$\ep$ such that $h(0)=0$. This implies $\hd\om_\ep=h(\ep)\hd\nu'_\ep$, which by
compairing with $\hd\om_\ep=f(\ep)\nu_\ep$ yields $h(\ep)\sim\ep$ and
$\nu_\ep\sim\hd\nu'_\ep$. Then
\be
\ep\sim\lan\nu^\ast_\ep|\chi_\ep\ran\sim\lan\hd\nu^{\prime\ast}_\ep|\chi_\ep\ran
=\lan\nu^{\prime\ast}_\ep|\hd\chi_\ep\ran=\ep\lan\nu^{\prime\ast}_\ep|\psi_\ep
\ran,\nonumber\ee
so that $\lan\nu^{\prime\ast}_\ep|\psi_\ep\ran\sim 1$, which contradicts
$\psi\in\CI$.\ $\Box$\vs{5mm}\\
{\sc Proposition 5.9}. $\HL$ for $p\geq 1$ are represented by elements
of the form $v\otimes\ti{v}_0\otimes v^{gh}$, or equivalently of
the form $v_0\otimes\ti{v}\otimes v^{gh}$, where $ v\in L$,
$\ti{v}\in \ti{L}$, $v_0$ is a primary highest weight vector w.r.t. $g$,
$\ti{v}_0$ is a primary highest weight vector w.r.t. $\ti{g}$ and
$v^{gh}$ satisfies $c(x)v^{gh}=0$, $x\in n_+$.\vs{3mm}\\
{\sc Proof.} Let $\hat{H}^p (g',L')$, $p\geq 1$ be non-zero. Then by Theorem 4.1
there exists $\om\in\CM$ such that $\pil(\om)\in\hat{H}^{-p} (g',L')$. We
have $\hd\om=\nu$
(Corollary 5.3) with  $\nu\in\CI$. It follows by Lemma 5.6 (iv) that
$\psi\in\CM$
with $\pil(\psi)\in\hat{H}^p (g',L')$, will satisfy  $\hd\psi=0$ in
$\CM$. We can now use the gradation $\ti{N}_{gr}$ introduced in
the previous section to decompose $\hd=\ti{d}_0+\ti{d}_{-1}$ and
use the homotopy operator $\ti{\ka}_0$ to successively eliminate
highest order terms of $\psi$ in this gradation. Since $p\geq 1$ we will
finally get an element of the form $v\otimes\ti{v}_0
\otimes v^{gh}$. The alternative form is found by using the
gradation $N_{gr}$.\ $\Box$\vs{5mm}\\
{\sc Proposition 5.10} . $\hat{H}^0 (g',M')$ are represented by
elements $v\otimes\ti{v}_0\otimes v^{gh}_0$, or equivalently by the elements
 $v_0\otimes\ti{v}\otimes v^{gh}_0$  ,
where $v,v_0$ and $\ti{v},\ti{v}_0$ are highest weight vectors w.r.t. $g$ and
$\ti{g}$, respectively, with $v_0$ and $\ti{v}_0$ being primary, and $v^{gh}_0$
is the ghost vacuum. Furthermore, the
weights $\mu$ and $\ti{\mu}$ of the primary highest weight vectors $v_0$
and $\ti{v}_0$,
respectively, satisfy $\mu+\ti{\mu}+\rho=0$.\vs{3mm}\\
{\sc Proof.} Let $\psi\in\hat{H}^0 (g',M')$. Then
$\hd\psi=0$ and we can use the gradation $\ti{N}_{gr}$ and the homotopy
operator as in the proof of Proposition 5.9 to conclude that since
$N^{gh}(\psi)=0$ we must have $\psi=v\otimes\ti{v}_0\otimes v^{gh}_0$. By
using the gradation $N_{gr}$ we get the alternative form. The condition on
the weights is a consequence
of $h^{tot}(v\otimes \ti{v}\otimes v^{gh}_0)=0$.\ $\Box$\vs{5mm}\\
{\sc Corollary 5.11} . $\hat{H}^0 (g',L')$ are represented
by elements of the form $v_0\otimes \ti{v}_0\otimes v^{gh}_0$. Furthermore, the
weights $\mu$ and $\ti{\mu}$ of the primary highest weight vectors $v_0$
and $\ti{v}_0$,
respectively, satisfy $\mu+\ti{\mu}+\rho=0$. \vs{3mm}\\
{\sc Proof.} Let $\psi\in\CM$ and $\pil(\psi)\in\hat{H}^0 (g',L')$. Assume first
$\hd\psi=0$ in $\CM$. Then the corollary follows directly from Proposition
5.10. Consider now
$\hd\psi=\nu\neq 0$, where $\pil(\nu)= 0$. Then by Lemma 5.6 (iv)
there exists $\om\in\CM$ such that $\hd\om=0$, $\om\not\in\CI$ and
$\langle\om^\ast|\psi\rangle\neq 0$. We may then apply Proposition 5.10 to
$\om$,
so that $\om$ is of the form claimed in the corollary. As $\langle \om^\ast
|\om\rangle\neq 0$, $\om$ is a singlet representation of the BRST
cohomology (cf.
Theorem 4.1) and, hence, $\psi$ and $\om$ yield equivalent elements in
$H^0(g',L')$.
\ $\Box$\vs{5mm}\\
{\sc Theorem 5.12}. A necessary and sufficient condition for $\hat{H}^{\pm p}
(g',L')$, $p\geq 1$, to be non-zero is either one of the following:\vs{2mm}\\
\hs{4mm}(i) There exists a vector $\nu\in\CI$ satisfying $\nu\not \in
\hat{C}(g',M^{\prime
(2)})$ and\\ \hs{4mm}$\nu\in\hat{H}^{-p+1}(g',M^{\prime (1)})$.\vs{1mm}\\
\hs{3mm}(ii) There exists a vector
$\chi\in\CI$ satisfying $\chi\not \in \hat{C}(g',M^{\prime (2)})$,\\
\hs{3mm}$N^{gh}(\chi)=p-1,\ \hd\chi=0$ and $\hd\chi_\ep\neq 0$.\vs{2mm}\\
In addition, $\mbox{dim}\left(\hat{H}^{-p+1}(g',M^{\prime (1)})\right)=
\mbox{dim}\left(\hat{H}^{\pm p} (g',L')\right)$, $p\geq 1$. \vs{3mm}\\ {\sc
Proof.} {\it
Necessary:} (i) follows by Corollary 5.3 and Lemma 5.6 (iii).  (ii) follows
from (i) together with Lemma 5.6 (ii) and (iii).\vs{1mm}\\ {\it Sufficient:}
(i) For $p>1$ we use Lemma 5.5. This also gives the last assertion of
dimensionalities for these cases. For $p=1$ we have $\nu
\in\hat{H}^{0}(g',M^{\prime (1)})$. We have two possibilities. Either $\nu\in
\hat{H}^{0}(g',M^{\prime })$ or $\nu=\hd\psi$ for some $\psi\in\CM$,
$\pil(\psi)\neq 0$. In
the first case we have $\hd\nu_\ep\neq 0$ from Proposition 5.10, so that we
get case (ii)
of the theorem, which is proved below. For the second possibility we use
Lemma 5.4.\vs{1mm}\\ (ii) $\hd\chi=0$ and $\hd\chi_\ep\neq 0$ implies,
using that $\hd$ is linear in the generators of $g'$,
$\hd\chi_\ep=\ep\psi_\ep$ for some $\psi_\ep$ satisfying $\lim_{\ep\rightarrow
0}\psi_\ep\neq 0$. Proposition 5.8
then gives $\pil(\psi)\in\HL$.

We finally prove the assertion concerning dimensionalities for the case $p=1$.
Assume first that there exist $\om_1,\ \om_2\in\CM$, with $\pil(\om_1),\
\pil(\om_2)\in
H^{-p}(g',L')$ and $\nu_1,\nu_2\in\hat{H}^{-p+1}(g\pri,M^{\prime(1)})$,
satisfying
$\nu_1=\hd\om_1,\nu_2=
\hd\om_2$ (which is necessary by Corollary 5.3), where $\nu_1=\nu_2$ mod
exact terms. This
implies
$\hd(\om_1-\om_2)=\nu_1-\nu_2=\hd(\ldots)$,  so that by Proposition 5.1,
$\pil(\om_1)=\pil(\om_2)$ mod exact terms. Consider the opposite case i.e.
two different
vectors $\nu_1$ and $\nu_2$ give the same element in $\hat{H}^{-1} (g',L')$.
Write $\hd\om_1=\nu_1$ and $\hd\om_2=\nu_2$. The requirement that
$\pil(\om_1)$ and
$\pil(\om_2)$ are equivalent elements in $\hat{H}^{\pm 1} (g',L')$ now implies
$\om_1=\om_2+\nu'+\hd(\ldots)$, where
$\nu'\in\CI$. Then  $\nu_1=\nu_2+\hd\nu'$.\ $\Box$\vs{5mm}\\
{\sc Corollary 5.13.} $\mbox{dim}\hat{H}^{\pm 1}(g',L'_{\mu \ti{\mu}})=1$ if
$l(\ti{\mu})-l(\mu)=1$ and $\mu$ and $-\ti{\mu}-\rho$ are on the same
$\rho$-centered Weyl
orbit, and  $\mbox{dim}\hat{H}^{\pm 1}(g',L'_{\mu \ti{\mu}})=0$
otherwise.\vs{5mm}\\
{\sc Proof.} By Proposition 5.10 and Theorem 3.11 we have
$\mbox{dim}\hat{H}^{0}(g',M^{\prime(1)})=1$ if
$l(\ti{\mu})-l(\mu)=0$ and $\mu$ and $-\ti{\mu}-\rho$ are on the same
$\rho$-centered Weyl orbit and $\mbox{dim}\hat{H}^{0}(g',M^{\prime(1)})=0$
otherwise. Then
$\mbox{dim}\hat{H}^{\pm 1}(g',L'_{\mu
\ti{\mu}})=$$\mbox{dim}\hat{H}^{0}(g',M^{\prime(1)})=1$ (Theorem 5.12).
With the help of
Proposition 5.10 we can easily construct
$\nu$ as in Theorem 5.12 (i), which gives the corollary.
$\Box$\vs{5mm}\\
{\sc Theorem 5.14.} $\hat{H}^{\pm p}(g',L'_{\mu \ti{\mu}})=0, p\geq
0,$ if $l(\ti{\mu})-l(\mu)\neq p$, or if $l(\ti{\mu})-l(\mu)=p$ and $\mu$ and
$-\ti{\mu}-\rho$ are not on the same $\rho$-centered Weyl orbit.\vs{5mm}\\
{\sc Proof.} The theorem is true for $p=0$ by Corollary 5.11 and for
$p=1$ by Corollary 5.13.
Assume the theorem to be true for $\hat{H}^{\pm q}(g',L'_{\mu \ti{\mu}})$,
$0\leq q\leq p-1$ and consider $q=p$.

Assume there exists $\om\in M_{\mu \ti{\mu}}^\prime$ such that
$\pi(\om)\in\hat{H}^{-p}(g',L'_{\mu \ti{\mu}})\neq 0$. Let $\mbox{grad}(\si)=s$
if $s$ is the largest integer for which $\si\in\hat{C}(g',M^{\prime (s)})$.
Then there
exists
$\nu\in\hat{H}^{-p+1}(g',M^{\prime (1)}_{\mu\ti{\mu}})$, $\mbox{grad}(\nu)=1$
(Theorem 5.12). Write $\nu=\nu_1+\nu_2+\ldots\nu_n$, where $\nu_i\in V_i$,
$i=1,2,\ldots,n$, $\mbox{grad}(\nu_i)=1$ and $V_i$ are Verma or BG modules
of primitive
weights
$(\mu_i,\ti{\mu}_i)$. We have $l(\ti{\mu}_i)-l(\mu_i)=l(\ti{\mu})-l(\mu)-1$
(Proposition 3.15). We may assume that $\nu$ cannot be written as a sum
$\nu^\prime+\nu^{\prime\prime}$ where $\nu^\prime,
\nu^{\prime\prime}\in\hat{H}^{-p+1}(g',M^{\prime (1)}_{\mu\ti{\mu}})$ and
unequal, $\mbox{grad}(\nu')=\mbox{grad}(\nu^{\prime\prime})=1$, as this would
yield two different elements in $\hat{H}^{-p}(g',L'_{\mu \ti{\mu}})$. If
$\hd\nu_i=0$ for some value of $i$, then $\nu_i=\hd(\ldots)$ (Proposition 5.10)
and clearly $\nu-\nu_i$ will correspond to the same element $\om$. Hence, we
may restrict to $\nu_i$ with $\hd\nu_i\neq 0, i=1,\ldots,n$.

Consider now the equation $\hd\nu=0$ using the gradation $N_{gr}$. Let
$\hat{\nu}_i$ be the highest order term of $\nu_i$ and
$N_{gr}(\nu_i)=N_i, i=1,\ldots,n$. Let $\hat{N}=\mbox{max}\{N_i\}_{i=1}^n$
and order so that $N_i=\hat{N}$ for $i\in \{1,2,\ldots,m\}$, $m\leq n$. Then
$d_0(\sum_{i=1}^m)\nu_i=0$. As $d_0\nu_i\in V_i$, this equation may only be
solved if
there exists at least one $V_j$ such that
$d_0\hat{\nu}_i\in V_i\cap V_j$. Let $\phi_{VMi}$ be the $g$-homomorphism
$V_i\stackrel{\phi_{VMi}}{\rightarrow}M_i\pri $, $i=1,\ldots, n$, where
$M_i$ are Verma
modules of the same primitive weight as $V_i$. $\phi_{VMi}$ exists for all $i$ (see the note
after Corollary 3.9). Then
$d_0\phi_{VMi}(\hat{\nu}_i)\in M_i\pri\cap M_j\pri$. This is only possible if
$d_0\phi_{VMi}(\hat{\nu}_i)\in M_i^{\prime (1)}$ for all $i$. This implies
$\hd\phi_{VMi}(\hat{\nu}_i)\in M_i^{\prime (1)}$ to highest order in
$N_{gr}$ and that there
exists $\eta_i\in M'_i$ such that
$\eta_i=\hd\phi_{VMi}(\nu_i)$. If there exists $\eta'_i\in M^{\prime
(1)}_i$ such that
$\eta=\hat{d}\eta'_i$ to leading order then
$\hd(\phi_{VMi}(\nu_i)-\eta'_i)$ to leading order,
which contradicts $\hd\phi_{VMi}(\nu_i)=\xi_i$, where $\xi_i$ is non-exact
in $M^{\prime
(1)}_i$. Hence,
$\eta_i\in\hat{H}^{-p+2}(g\pri,M^{\prime (1)}_{\mu_i\ti{\mu}_i})$ to
highest order.
Theorem 5.12 now asserts that $\pi_{L_i}\phi_{VMi}(\hat{\nu}_i)\in
\hat{H}^{-p+1}(g',L'_i)$ to highest order. The induction hypothesis implies
$l(\ti{\mu}_i)-l(\mu_i)=p-1$ and that $\mu_i$ and $-\ti{\mu}_i-\rho$ lie on the
same $\rho$-centered Weyl orbit. Then $l(\ti{\mu})-l(\mu)=p$, $\mu$ and
$-\ti{\mu}-\rho$
lie on the same Weyl orbit (Theorem 3.11 and Proposition 3.15).\
$\Box$\vs{5mm}\\
{\sc Theorem 5.15.} Let $\mu,\ti{\mu}\in\Ga_w$ be such that $\mu+\rho/2$ and
$\ti{\mu}+\rho/2$ are regular and $\mu$ and $-\ti{\mu}-\rho$ are on the same
$\rho-$centered Weyl orbit. Then
$\hat{H}^{\pm p}(g',L_{\mu\ti{\mu}})\neq 0$, where
$p=l(\ti{\mu})-l(\mu)\geq 0$. \vs{5mm}\\
{\sc Proof.} For $p=0$ the theorem is given by Corollary 5.11 (cf. the
proof of Theorem
5.14, where it is shown that $\mu+\ti{\mu}+\rho=0$ implies
$l(\ti{\mu})-l(\mu)=0$). For $p=1$ the theorem follows from Corollary 5.13. We
proceed by induction on $p$. Assume the theorem to be true for
$0\leq l(\ti{\mu})-l(\mu)\leq p-1$. We will also assume the following to hold to
this order of $p$. Let $\om\in M_{\mu\ti{\mu}}$ such that
$\pi_L(\om)\in\hat{H}^{-q}(g',L_{\mu\ti{\mu}})$ for $0\leq q\leq p-1$. We then
assume $\hd\om=\nu_1+\nu_2+\ldots+\nu_n$ with $\nu_i\in M_{\mu_i\ti{\mu}_i}$
and $\mbox{grad}(\nu_i)=1, i=1,\ldots , n$ (with grad(...) defined as in
Theorem 5.14).
This assumption clearly holds for
$p=1$.

We now consider $\mu, \ti{\mu}$ such that $l(\ti{\mu})-l(\mu)=p\geq 2$ with
$\ti{\mu}+\rho/2$ and $\mu+\rho/2$ being regular. Introduce the following
notation. For
the Verma module $M_{\mu\ti{\mu}}$ we let $M_1,\ldots,M_n$ denote all submodules
 such that
$M_i\subset M^{(1)}_{\mu\ti{\mu}}$, $M_i\not\subset M^{(2)}_{\mu\ti{\mu}}$,
$i=1,\ldots, n$. Denote by
$(\mu_1\ti{\mu}_1),\ldots,(\mu_n\ti{\mu}_n)$ their respective highest weights.
Let $\phi_i$
be a non-zero element of
$\mbox{Hom}_g(M_i,M_{\mu\ti{\mu}})$, $i=1,\ldots, n$. Let $M_{i_1\ldots
i_k}=M_{i_1}\bigcup\ldots\bigcup M_{i_k}$, $i_1,\ldots,i_k=1,\ldots ,n$ and
$\phi_{i_1\ldots i_k}$ be a non-zero element of $\mbox{Hom}_g(M_{i_1\ldots
i_k},M_{\mu\ti{\mu}})$.
Consider now $\om_1\in M_1$ with
$\pi_L(\om_1)\in\hat{H}^{-p+1}(g',L_{\mu_1\ti{\mu}_1})$. By Theorem 3.11
and the induction
hypothesis $\om_1$ exists. Then $\hd\om_1=\nu_1+\ldots+\nu_s$, where
$\nu_i\in M_{1,i}\subset M^{(1)}_1$ (induction hypothesis). As
grad$(\nu_i)$=1 we have
grad$(\phi_i(\nu_i))$=2. Therefore, there will exist a union of Verma modules,
$M_{2\ldots k}$ say, such that
$\phi_1(\nu_1+\ldots+\nu_s)=\phi_{2\ldots k}(\nu\pri_1+\ldots+\nu\pri_t)$,
$\nu\pri_1+\ldots+\nu\pri_t\in M_{2\ldots k}$. By Lemma 3.17 $M_{2\ldots
k}$ is non-zero
and different from $M_1$.
 Thus,
$\phi_1(\nu_1+\ldots+\nu_s)$ may either be viewed as originating from an element
$\nu_1+\ldots+\nu_s$ in
$M_1$ or from an element $\nu\pri_1+\ldots+\nu\pri_t$ in
$M_{2\ldots k}$. As $\hd (\nu\pri_1+\ldots+\nu\pri_t)=0$, there exists an
element
$\om_{2\ldots k}\in M_{2\ldots k}$ such that
$\nu\pri_1+\ldots+\nu\pri_t=\hd\om_{2\ldots k}$ (Proposition 5.1).
From this it follows that $\hd(\phi_{2\ldots k}(\om_{2\ldots
k})-\phi_1(\om_1))=0$.
Define $\xi=\phi_{2\ldots k}(\om_{2\ldots k})-\phi_1(\om_1)$, which must be
non-zero as
$M_1\neq M_{2\ldots k}$. We now prove that
$\xi$ is a non-trivial element of
$\hat{H}^{-p+1}(g',M^{\prime(1)})$, which by Theorem 5.12 proves our
assertions for
$l(\ti{\mu})-l(\mu)=p$ (including the additional induction assumption).

Assume the contrary, $\xi=\hd\eta$ with $\eta\in M^{(1)}_{\mu\ti{\mu}}$.
 Let $\xi=\xi_1+\ldots\xi_k$, where $\xi_i\in M_i$,
$i=1,\ldots, k$. By construction $\hd\xi_i\in M^{(2)}_{\mu\ti{\mu}}$,
$i=1,\ldots, k$. If $\hd\xi_i=0$ then the corresponding Verma module $M_i$
may be
deleted from $M_{2\ldots k}$ without affecting the construction of $\xi$ (as
$\hd\xi_i=0$ implies $\xi_i=\hd(\ldots)$). Hence, we may consider $\hd\xi_i\neq
0$, $i=1,\ldots, k$. To highest order in the gradation $\ti{N}_{gr}$ the
equation
$\xi=\hd\eta$ yields
$\xi^{(N)}=\xi_i^{(N)}+\ldots+\xi^{(N)}_k=\ti{d}_0\eta^{(N)}$, where
$\xi^{(N)}$ and $\eta^{(N)}$ are the leading terms in $\xi$ and $\eta$,
respectively, and
$\xi_i^{(N)}$ denotes the N'th order term of
$\xi_i$ (which is non-zero for at least one value of $i$). Generally,
$\ti{d}_0\ga\in V$ only if $\ga\in V$ for any Verma or BG module V. Therefore,
$\eta^{(N)}=\eta_1^{(N)}+\ldots \eta_k^{(N)}$ and
$\xi_i^{(N)}=\ti{d}_0\eta_i^{(N)}$
$\eta_i^{(N)}\in M_i$ $i=1,\ldots ,k$. This implies $\ti{d}_0\xi^{(N)}=0$
and in turn
$\hd\xi_i=0$, which is a contradiction.\
$\Box$\vs{5mm}\\
{\sc Remark 1:} Results similar to Theorem 5.15 may be obtained for weights
$\mu+\rho/2$
and $\ti{\mu}+\rho/2$ being singular, provided the corresponding Verma
modules satisfy
the multiplicity condition of Lemma 3.17. It is clear, however, that this
generalization does not hold for all singular cases.\vs{5mm}\\
{\sc Remark 2:} The proof of Theorem 5.15 provides also an explicit method
for finding the
elements of the cohomology for negative ghost numbers. It is the same
method as was
presented in ref \cite{HR2}.\vs{5mm}\\
{\bf Acknowledgement:} I would like to thank Henric Rhedin for stimulating
discussions during the progression of this work.

\end{document}